\documentclass[aps,twocolumn,prx,preprintnumbers,showpacs,amsmath,amssymb,superscriptaddress]{revtex4-1}
\usepackage[pdftex, colorlinks, citecolor=blue]{hyperref}   
\usepackage{graphicx}
\usepackage{dcolumn}
\usepackage{threeparttable}
\usepackage{multirow}
\usepackage{booktabs}
\usepackage{txfonts}
\usepackage{xcolor}
\usepackage{bm}
\usepackage{amssymb}
\usepackage{amsmath}
\usepackage{latexsym}
\usepackage{epsfig}
\usepackage{amsbsy}
\usepackage{array}
\usepackage{tabularx}
\usepackage{esvect} 
\usepackage{extarrows}
\usepackage{float}
\usepackage[british]{babel}
\usepackage{longtable}
\usepackage{makecell}

\begin{document}
	
\title{Comprehensive \emph{ab initio} study of effects of alloying elements on generalized stacking fault energies of Ni and Ni$_3$Al}
	
\author{Heyu Zhu}
\affiliation{%
Shenyang National Laboratory for Materials Science, Institute of Metal Research, Chinese Academy of Sciences, 110016 Shenyang, China.
}%
\affiliation{%
School of Materials Science and Engineering, University of Science and Technology of China, 110016 Shenyang, China.
}%
	
\author{Jiantao Wang}
\affiliation{%
Shenyang National Laboratory for Materials Science, Institute of Metal Research, Chinese Academy of Sciences, 110016 Shenyang, China.
}%
\affiliation{%
School of Materials Science and Engineering, University of Science and Technology of China, 110016 Shenyang, China.
}%
	
\author{Yun Chen}
\affiliation{%
Shenyang National Laboratory for Materials Science, Institute of Metal Research, Chinese Academy of Sciences, 110016 Shenyang, China.
}%

\author{Mingfeng Liu}
\affiliation{%
Shenyang National Laboratory for Materials Science, Institute of Metal Research, Chinese Academy of Sciences, 110016 Shenyang, China.
}%
\affiliation{%
School of Materials Science and Engineering, University of Science and Technology of China, 110016 Shenyang, China.
}%

\author{Hui Ma}
\affiliation{%
Shenyang National Laboratory for Materials Science, Institute of Metal Research, Chinese Academy of Sciences, 110016 Shenyang, China.
}%

\author{Yan Sun}
\affiliation{%
Shenyang National Laboratory for Materials Science, Institute of Metal Research, Chinese Academy of Sciences, 110016 Shenyang, China.
}%
	
\author{Peitao Liu}%
\email{ptliu@imr.ac.cn}
\affiliation{%
Shenyang National Laboratory for Materials Science, Institute of Metal Research, Chinese Academy of Sciences, 110016 Shenyang, China.
}%
	
\author{Xing-Qiu Chen}%
\email{xingqiu.chen@imr.ac.cn}
\affiliation{%
Shenyang National Laboratory for Materials Science, Institute of Metal Research, Chinese Academy of Sciences, 110016 Shenyang, China.
}%
	
\begin{abstract}
Excellent high-temperature mechanical properties
of Ni-based single crystal superalloys (NSCSs) are attributed to the yield strength anomaly
of Ni$_{3}$Al that is intimately related to generalized stacking fault energies (GSFEs).
Therefore, clarifying the effects of alloying elements on the GSFEs is of great significance for alloys design.
Here, by means of \emph{ab initio} density functional theory calculations,
we systematically calculated the GSFEs of different slip systems of Ni and Ni$_{3}$Al
without and with alloying elements using the alias shear method.
We obtained that for Ni, except for magnetic elements Mn, Fe, and Co,
most of alloying elements decrease the unstable stacking fault energy
($\gamma_{usf}$) of the $[01\bar{1}](111)$ and $[11\bar{2}](111)$ slip systems
and also decrease the stable stacking fault energy ($\gamma_{sf}$) of the $[11\bar{2}](111)$ slip system.
Interestingly, the reduction effects exhibit a strong correlation with the inverse of atom radii.
For Ni$_{3}$Al, most of alloying elements in groups IIIB-VIIB show a strong Al site preference.
Except for Mn and Fe, the elements in groups VB-VIIB and the first column of group VIII increase the values of
$\gamma_{usf}$ of different slip systems of Ni$_{3}$Al,
which makes the slip deformation and dislocation emits difficult.
On the other hand, the elements in groups IIIB-VIIB also increase the value of $\gamma_{sf}$,
and thus reduce the stability of the antiphase boundary, complex stacking fault and
superlattice intrinsic stacking fault of Ni$_{3}$Al.
We found that Re is an excellent strengthening alloying element that significantly
increases the slip barrier of the tailing slip process for Ni,
and also enhances the slip barrier of the leading slip process of three slip systems for Ni$_{3}$Al.
W and Mo exhibit similar effects as Re.
We predicted that Os, Ru, and Ir are good strengthening alloying elements as well,
since they show the strengthening effects on both the leading and tailing slip process for Ni and Ni$_{3}$Al.
This work established an exhaustive dictionary of the effects of various alloying elements on the GSFEs of both Ni and Ni$_3$Al phases,
which would aid to guide the design of next-generation high-performance NSCSs.
\end{abstract}
	
\maketitle

\section{Introduction}

Nickel-based single crystal superalloys (NSCSs) exhibit excellent mechanical properties
due to the formation of a large volume fraction of ordered Ni$_{3}$Al precipitates
that are coherently embedded in the matrix of the Ni phase~\cite{2006_Reed, 2015_Royce, 2006_Pollock, 1999_Caron}.
These ordered precipitates lead to order strengthening and anomalous temperature dependence of yield strength,
thereby resulting in extraordinarily high strength and creep resistance at elevated temperatures~\cite{doi10.2514/1.18239}.

As an intrinsic property of the materials with the $L1_{2}$ structure~\cite{KARNTHALER1996547, UMAKOSHI1984449},
the yield strength anomaly (YSA) is intimately connected to the stacking faults (SFs) in the (111) plane of Ni$_{3}$Al,
including the antiphase boundary (APB),
the superlattice intrinsic stacking fault (SISF),
and the complex stacking fault (CSF)~\cite{PhysRevB.101.024102}.
For example, it was found that the
thermally activated cross slip of screw dislocations from the \{111\} primary slip plane to the \{001\} cross slip plane
is responsible for the YSA occurrence~\cite{doi10.1080/095008399176544, doi10.1080/02670836.2016.1215961},
since the cross slip remains locked in Kear-Wilsdorf (KW) configurations~\cite{doi10.1080/02670836.2016.1215961}.
The driving force to form the KW locks was found to be positively correlated with
the APB energy ($\gamma_{\rm{APB}}$)~\cite{abate_nanoscopy_2016} according to the Paidar-Pope-Vitek model~\cite{PAIDAR1984435}.
Moreover, the KW locks tend to form when the CSF energy ($\gamma_{\rm{CSF}}$) is low~\cite{KARNTHALER1996547, chen_modeling_2022}.
Furthermore, the dislocation ribbons of overall Burgers vector $[\bar{1}12]$ shearing the Ni$_{3}$Al precipitates
can also affect the YSA~\cite{doi10.1080/01418618608242846, RAE20071067, MATAN19991549, kear_stacking_1970}.
Specifically, at intermediate temperatures and high stresses,
the $[1\bar{1}0]$ dislocations first form in the matrix of the Ni phase
and then decompose at the Ni/Ni$_{3}$Al interface in the end of the first stage of creep or at the beginning of the second stage of creep~\cite{RAE20071067}.
The decomposition reaction can be described by
$1/2[011] + 1/2[\bar{1}01] \rightarrow 1/3[\bar{1}12] + 1/6[\bar{1}12]$~\cite{RAE20071067}.
The leading partial dislocation $1/3[\bar{1}12]$ enters the Ni$_{3}$Al phase and creates a SISF,
while the tailing partial dislocation $1/6[\bar{1}12]$  remains at the Ni/Ni$_{3}$Al interface~\cite{RAE20071067}.
The formation of the $[\bar{1}12]$ dislocation ribbons was first observed by Leverant and Kear using transmission electron microscopy~\cite{Leverant1973}
and later confirmed by other studies~\cite{MATAN19991549,KNOWLES200388, RAE20071067,BREIDI201897}.
The SISF has a lower energy as compared to other SF configurations and
was thought to be the dominant SF that drives the shearing of the Ni$_{3}$Al phase
under high stress and intermediate-temperature creep conditions~\cite{RAE20071067, RAE2001125, KOVARIK2009839, VISWANATHAN20053041, MRYASOV20024545}.
Besides, the formation of high-density SFs can promote the accumulation of partial dislocations,
thereby enhancing both the ductility and fracture toughness but without compromising high strengths.
Considering the significant effects of the SFs in improving the creep behavior of NSCSs,
controllable tuning the SF energy by alloying would be highly desirable for the design of high-performance NSCSs.

Experimentally, it is tough to obtain an accurate value of the SF energy of NSCSs,
because the separation distance between partial dislocations in electron microscope images
is too small to identify~\cite{PhysRevB.75.224105, VITOS20063821}.
Based on the weak-beam method of electron microscopy,
the derived SF energies of the Ni phase normally lie in the range of 120$\sim$130 $mJ/m^2$~\cite{doi10.1080/14786437708232942, decampos2008}.
It is well known that to acquire high-performance NSCSs, more than ten alloying elements have been added to the NSCSs so far,
such as Ti, V, Cr, Co, Zr, Nb, Mo, Ru, Hf, Ta, W and Re.
For instance, it was experimentally found that Re and Co can reduce the SF energy of Ni~\cite{MA20075802}.
Similar effects were obtained by other alloying elements such as Co, Cr, Mo, Ti and W~\cite{gallagher_influence_1970,XIE1982483}.

In contrast to experiments, theoretical calculations in particular
the \emph{ab initio} calculations based on the density functional theory (DFT)
have demonstrated their increasing power in alloys design, providing important
complementary perspectives in guiding the experimental
studies~\cite{10.3389/fmats.2020.00290, IKEDA2019464, Curtarolo2013, Hart2021}.
As early as in the year of 1968, Vitek proposed the generalized stacking fault energy (GSFE) model
to explain the effect of SF energy on shear deformation~\cite{doi10.1080/14786436808227500}.
In this model, the local minimum and maximum of the GSFE along the pathway of slip system are regarded as
stable stacking fault energy ($\gamma_{sf}$) and unstable stacking fault energy ($\gamma_{usf}$), respectively.
In general, the low $\gamma_{sf}$ can lead to
a larger width of stacking faults~\cite{LI20094988},
a higher strain-hardening coefficient~\cite{PIERCE2015178},
a higher twinnability~\cite{TADMOR20042507, CAHOON200956},
a lower twinning stress~\cite{SARMA20107624}
and a lower occurrence of cross-slip (or climb)~\cite{decampos2008, doi10.1063/1.3585786}
as well as a lower steady-state creep rate~\cite{MOHAMED1974779, ARGON1981293, YU20095914}.
The GSFE model combined with \emph{ab initio} calculations has
been widely adopted to study the effects of the alloying elements on the SF energies of Ni and
Ni$_{3}$Al~\cite{Shang_2012, Shang_2012_2, doi10.1063/1.2051793, YU20095914,
YU201238, EURICH201587, YANG2020109682, HU2020155799, XIA2022104183, ZHAO2022110990}.
For instance, Yu \emph{et al.}~\cite{YU20095914} found that Mo, Re and W remarkably
decrease the $\gamma_{sf}$ of Ni due to the $d$-$d$ orbitals hybridizations between solute and Ni.
Shang \emph{et al.}~\cite{Shang_2012, Shang_2012_2}
obtained that almost all the alloying elements decrease the $\gamma_{sf}$ of Ni,
with the effect being more pronounced as the alloying element is far from Ni in the periodic table.

\begin{figure}
\begin{center}
\includegraphics[width=0.48\textwidth]{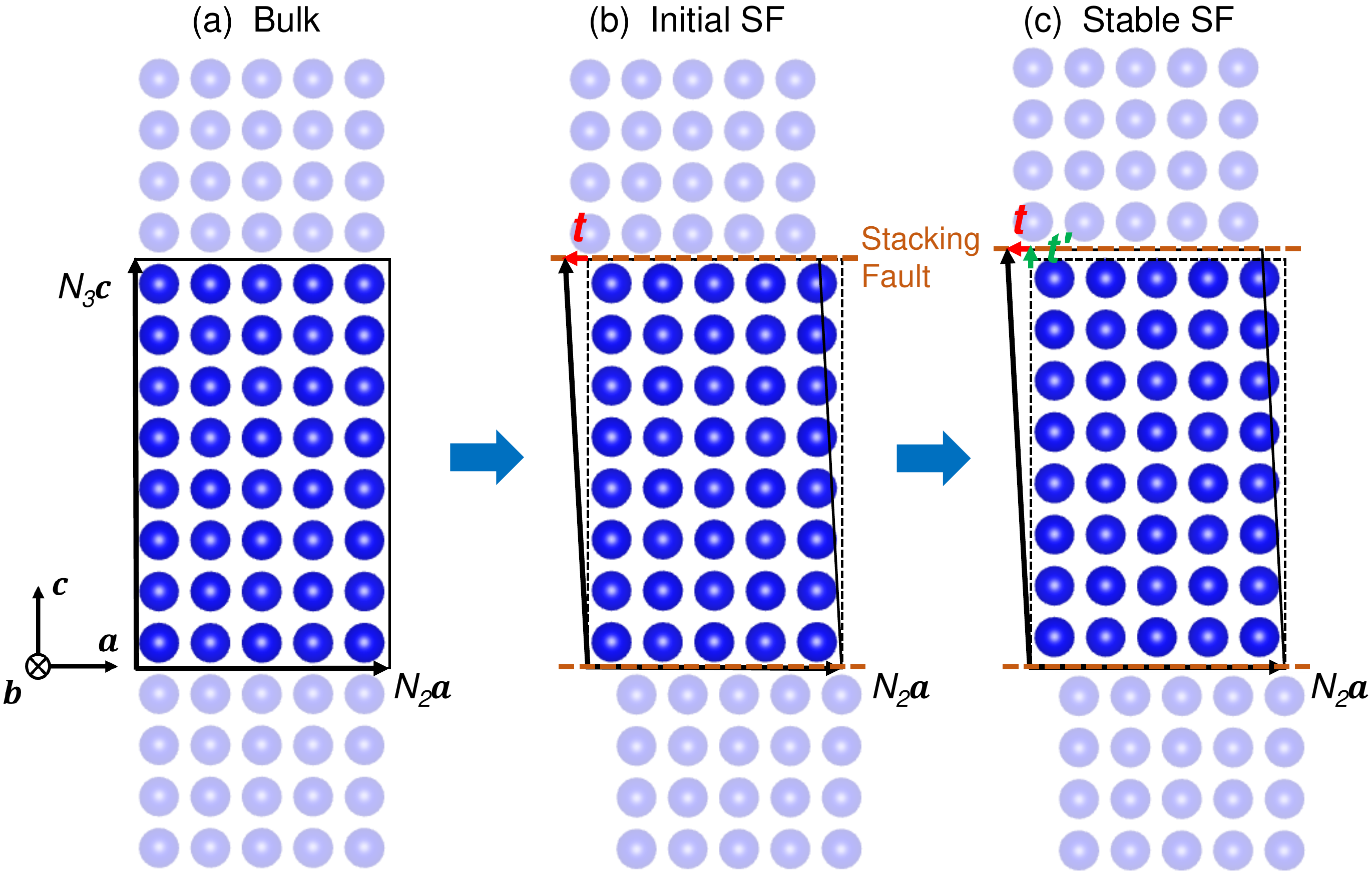}
\end{center}
\caption{Sketch of stacking fault (brown dashed lines) modeled by the alias shear method.
The lattice containing dark blue balls indicates the supercell used for calculating the GSFE.
(a) Undistorted supercell viewed in the $ac$ plane.
(b) Initial supercell with an ideal stacking fault,
which is created by $N_3 \boldsymbol{c}\rightarrow N_3 \boldsymbol{c} + \boldsymbol{t}$
with $\boldsymbol{t}$ being a displacement vector along the slip direction (here along $\boldsymbol{a}$).
Note that the atomic positions remain fixed in their initial cartesian coordinates.
(c) Supercell after structural relaxations along the $c$ direction only.
Here, $\boldsymbol{t^\prime}$ denotes the change of the lattice vector along $\boldsymbol{c}$.
}
\label{fig:alias}
\end{figure}

Although the effects of alloying elements on the SF energies of the Ni phase have been extensively studied~\cite{Shang_2012, Shang_2012_2, doi10.1063/1.2051793, YU20095914, YU201238, EURICH201587, YANG2020109682, HU2020155799, XIA2022104183, ZHAO2022110990},
they are normally done on a case-by-case basis and a systematic and thorough investigation of
typical alloying elements in the periodic table is still lacking,
in particular for the Ni$_3$Al phase
for which only a few alloying elements' effects on the SF energies have been reported~\cite{ZHAO2022110990}.
The contribution of this work is to establish a general and exhaustive dictionary
of the effects of various alloying elements on the GSFEs of different slip systems of both Ni and Ni$_3$Al phases.
This not only allows us to identify good strengthening alloying elements,
but also enables to determine alternative alloying elements that exhibit similar strengthening effects.
This is what the materials scientists desire and eventually would help to guide the design of the next-generation high-performance NSCSs.

This work is organized as follows.
First, the computational model used to compute the GSFEs
for various slip systems of the (111) plane of Ni and Ni$_{3}$Al is introduced.
Then, the GSFEs of the (111) plane of pure Ni and Ni$_{3}$Al
as well as their corresponding alloys are calculated.
The focus is on elucidating the effects of the considered 29 alloying elements
on the values of $\gamma_{sf}$ and $\gamma_{usf}$ of different slip systems.
Finally, conclusions are drawn.

\section{Methods and computational details}

All first-principles calculations were performed
using the Vienna \emph{ab initio} simulation package (VASP)~\cite{Kresse1996Efficiency, PhysRevB.54.11169}.
The generalized gradient approximation~\cite{PhysRevLett.77.3865}
parameterized by Perdew-Burke-Ernzerhof (PBE) was employed for the exchange-correlation functional.
The plane wave cutoff energy was set to 420 eV and the Brillouin zone was sampled
by a $\Gamma$-centered $k$-point grid
with the smallest allowed spacing between $k$ points of  0.16 $\AA^{-1}$,
The convergence criteria for the total energy and ionic forces were set to 10$^{-6}$ eV
and 0.01 eV/$\AA$, respectively.
The first-order Methfessel-Paxton method~\cite{PhysRevB.40.3616}
with a smearing width of 0.18 eV was used for structure relaxations,
whereas the Bl\"{o}chl-corrected tetrahedron method~\cite{PhysRevB.49.16223}
was used to obtain more accurate total energies. For all the calculations, spin polarization was considered.

To calculate the GSFEs, two approaches for modeling the SFs have been proposed.
The first one is the slab shear method~\cite{doi10.1063/1.2051793, DATTA2009124, HAN2011693}
and the other one is the alias shear method~\cite{Shang_2012, Shang_2012_2, doi10.1126/science.1076652, PhysRevB.79.224103}.
For the slab shear method, the SF is modeled by equally splitting a supercell into two slabs,
where the atoms in the bottom slab are fixed, while the atoms in the upper slab undergo a certain displacement along the slip direction.
By contrast, in the alias shear method the SF is complemented by an alias shear deformation of a periodic supercell lattice,
whereas the atomic positions are still represented by the initial cartesian coordinates.
As compared to the slab shear method,
the required supercell for modeling the SF is reduced by a factor of two in the alias shear method,
significantly decreasing the computational cost.
For this reason, the alias shear method was employed to calculate the GSFEs throughout the work.

Figure~\ref{fig:alias} sketches the SF model using the alias shear method.
First, a supercell with lattice vectors of
$N_{1}\boldsymbol{a}$, $N_{2}\boldsymbol{b}$ and $N_{3}\boldsymbol{c}$ was built [Fig.~\ref{fig:alias}(a)].
Then, a small vector $\boldsymbol{t}$ along the slip direction (here $\boldsymbol{a}$) was
introduced on the lattice vector $N_{3}\boldsymbol{c}$,
which constructs a slip plane in the $ab$ plane [see Fig.~\ref{fig:alias}(b)].
Finally, the supercell (including both lattice vectors and atomic positions) was allowed to relax only along the $c$ direction,
giving rise to a net vector of $\boldsymbol{t^\prime}$ perpendicular to the slip plane.
Here, we employed a 96-atom orthorhombic supercell with the lattice vectors of [11$\bar{2}$]$a_{0}$, 2[$\bar{1}$10]$a_{0}$, and 2[111]$a_{0}$
with $a_{0}$  being the lattice constant of conventional fcc unit cell.
The length of the lattice vector along the [111] direction ensures to eliminate interactions between periodic stacking faults planes.
The GSFE was calculated by
\begin{align}\label{eq:gamma_GSFE}
\gamma_{\rm{GSF}} = (E_{\rm{GSF}}-E_{0})/A,
\end{align}
where $E_{\rm{GSF}}$ and $E_{0}$ represent the total energies of the supercell with or without the SF, respectively,
and $A$ is the area of the slip plane within the supercell.
In order to calculate the $\gamma_{\rm{GSF}}$ with an alloying element,
for Ni$_{95}$X we substituted one of Ni atoms in the GSF plane with an alloying element.
Meanwhile, we also replaced a Ni atom in the GSF-free supercell.
Since the solute is placed in the stacking fault plane, this translates to a planar solute concentration of 6.25 at.\%.
Similar procedure was performed for Ni$_{72}$Al$_{23}$X, but here only
the substitution of Al atom in the GSF plane was considered due to its overall small normalized transfer energy
as compared to that when submitting Ni atom~\cite{PhysRevB.55.856, JIANG2006433, WU2012436}.

\begin{figure}
	\begin{center}
		\includegraphics[width=0.45\textwidth]{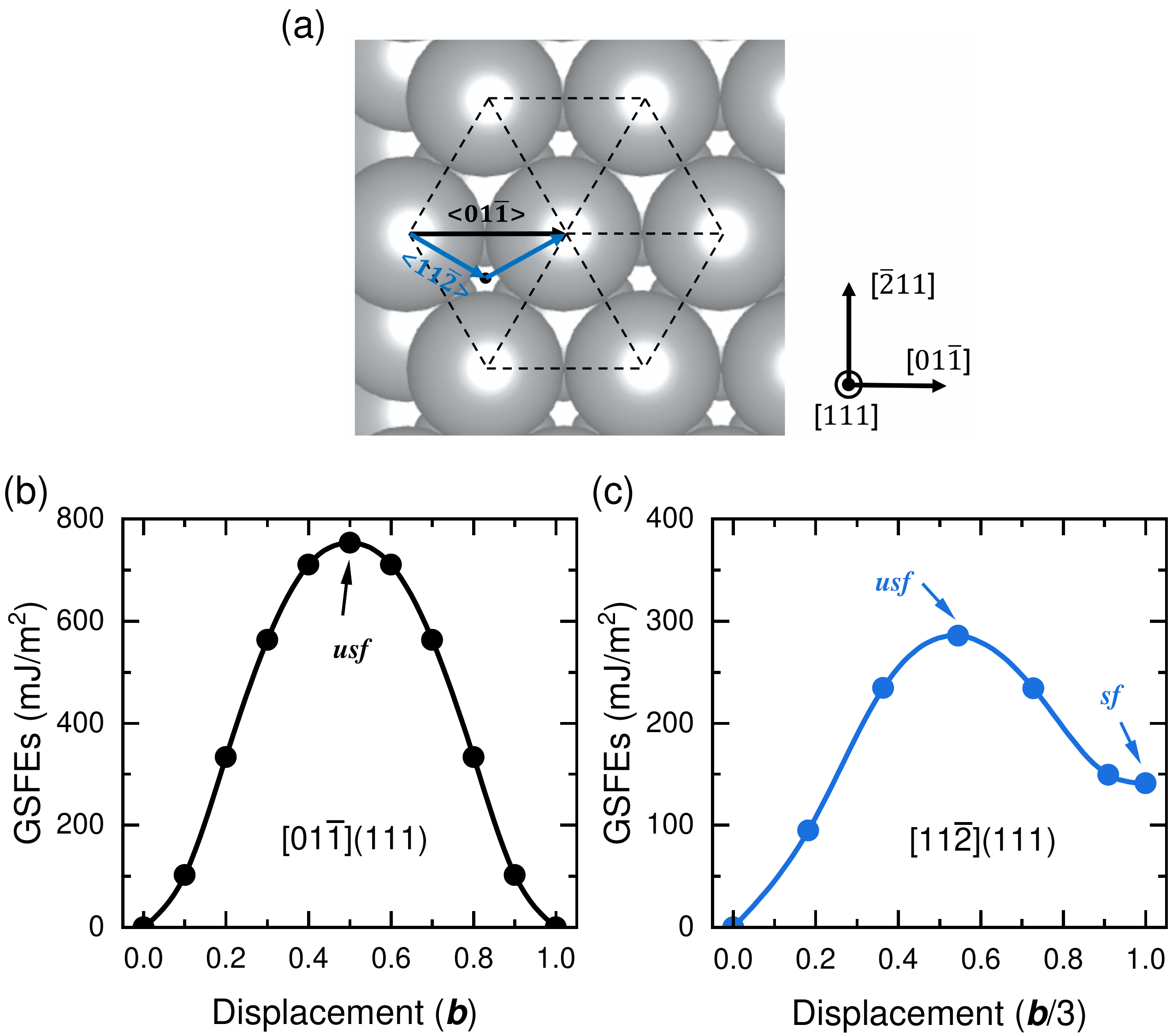}
	\end{center}
	\caption{(a) Sketch of two slip systems of pure Ni,
		i.e., $[01\bar{1}](111)$ and 1/3$[11\bar{2}](111)$.
		The corresponding GSFEs are shown in (b) and (c).
	}
	\label{fig:pure_ni}
\end{figure}

\section{RESULTS AND DISCUSSION}

\subsection{The GSFEs of pure Ni and Ni$_{3}$Al}

Let us start from the pure Ni system.
For the pure Ni, there are two typical slip systems, i.e., $[01\bar{1}](111)$ and $[11\bar{2}](111)$ [see Fig.~\ref{fig:pure_ni}(a)].
The corresponding calculated GSFEs are given in Table~\ref{tab:pure_GSFEs} and further plotted in Figs.~\ref{fig:pure_ni}(b) and (c).
The $\gamma_{usf}$ along the $[01\bar{1}]$ direction is computed to be 755 $mJ/m^2$,
which appears at $\boldsymbol{b}$/2 burgers vector,
whereas the calculated $\gamma_{usf}$ along the $[11\bar{2}]$ direction appearing at $\boldsymbol{b}/6$ burgers vector
is much smaller (285 $mJ/m^2$).
This indicates that launching the $[11\bar{2}](111)$ slip system would
be easier as compared to the $[01\bar{1}](111)$ slip system.
This is consistent with the Rice criterion $D = 0.3 \gamma_{surf}/\gamma_{usf}$
[$\gamma_{surf}$ is the surface energy, 1906 $mJ/m^2$ for Ni(111)]
showing that the larger the $D$ value is (equivalently smaller $\gamma_{usf}$), the more ductile is the slip system~\cite{RICE1992239}.
Our calculated $\gamma_{sf}$ and $\gamma_{usf}$ are overall in line with the literature theoretical and experimental data (see Table~\ref{tab:pure_GSFEs}).

\begin{table*}
\renewcommand\arraystretch{1.2}
\caption{Calculated stable stacking fault energies ($\gamma_{sf}$) and unstable stacking fault energies ($\gamma_{usf}$)
of different slip systems in Ni and Ni$_{3}$Al, which are compared to other literature data.}\label{tab:pure_GSFEs}
\begin{ruledtabular}
\begin{tabular}{llll}
Slip system  &  $\gamma_{sf}$ ($mJ/m^{2}$)  &  $\gamma_{usf}$ ($mJ/m^{2}$)	 & Notes and references  \\ \hline
Ni: $[01\bar{1}](111)$  &      & 755  & Calc., this work, alias shear  \\
                        &      & 783  & Calc., DFT, alias shear~\cite{HU2020155799}   \\
\hline
Ni: $[11\bar{2}](111)$   & 141  & 285  & Calc., this work, alias shear  \\
                        & 160  & 283  & Calc., DFT, alias shear~\cite{HU2020155799}   \\
                        & 149  & 273  & Calc., DFT, slab shear~\cite{XIA2022104183}   \\
                        & 129  & 278  & Calc., DFT, slab shear~\cite{YANG2020109682}   \\
    & 120 $\sim$ 130  &      & Expt., weak-beam TEM images~\cite{doi10.1080/14786437708232942} \\
\hline
Ni$_{3}$Al: $[01\bar{1}](111)$  & 244  & 819  & Calc., this work, alias shear  \\
(APB)                          & 259  & 830  & Calc., DFT, alias shear~\cite{HU2020155799}   \\
                               & 198  & 791  & Calc., DFT, slab shear~\cite{XIA2022104183}   \\
                               & 180  & 778  & Calc., DFT, slab shear~\cite{YU201238}   \\
                               & 210  &      & Calc., Peierls-Nabarro model~\cite{MRYASOV20024545}   \\
            &175$\pm$15 &      & Expt., weak-beam TEM images~\cite{KARNTHALER1996547} \\
\hline
Ni$_{3}$Al: $[\bar{1}2\bar{1}](111)$ & 214  & 257  & Calc., this work, alias shear  \\
(CSF)                               & 249  &      & Calc., DFT, alias shear~\cite{HU2020155799}   \\
                                    & 208  & 227  & Calc., DFT, slab shear~\cite{XIA2022104183}   \\
                                    & 205  & 254  & Calc., DFT, slab shear~\cite{YU201238}   \\
                                    & 225  &      & Calc., Peierls-Nabarro model~\cite{MRYASOV20024545}   \\
                 &235$\pm$45  &      & Expt., weak-beam TEM images~\cite{KARNTHALER1996547} \\
\hline
Ni$_{3}$Al: $[11\bar{2}](111)$  & 68   & 1339 & Calc., this work, alias shear  \\
(SISF)                         & 47   & 1421 & Calc., DFT, alias shear~\cite{HU2020155799}   \\
                               & 21   & 1332 & Calc., DFT, slab shear~\cite{XIA2022104183}   \\
                               & 75   & 1368 & Calc., DFT, slab shear~\cite{YU201238}   \\
                               & 80  &      & Calc., Peierls-Nabarro model~\cite{MRYASOV20024545}   \\
            &6$\pm$0.5  &      & Expt., weak-beam TEM images~\cite{KARNTHALER1996547} \\
                        &35  &      & Expt., weak-beam TEM images~\cite{KNOWLES200388} \\
\end{tabular}
\end{ruledtabular}
\end{table*}

\begin{figure}
\begin{center}
\includegraphics[width=0.45\textwidth]{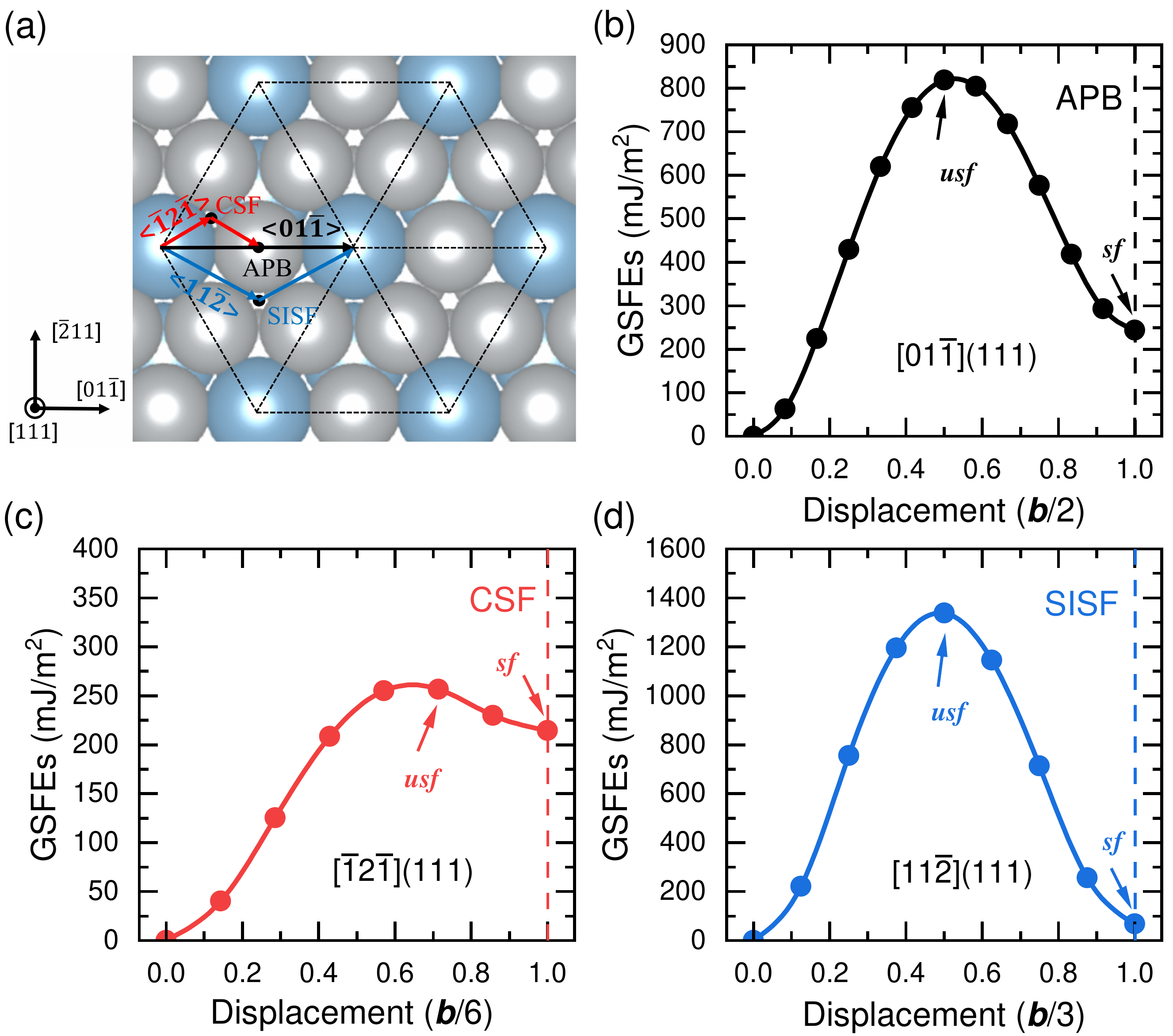}
\end{center}
\caption{(a) Sketch of three slip systems of pure Ni$_{3}$Al,
i.e., $[01\bar{1}](111)$, $[\bar{1}2\bar{1}](111)$ and $[11\bar{2}](111)$.
The corresponding GSFEs are shown in (b), (c) and (d).
}
\label{fig:pure_ni3al}
\end{figure}

Next, we move to the pure Ni$_{3}$Al system.
For the pure Ni$_{3}$Al, three inequivalent slip systems exist on the (111) plane,
i.e., $[01\bar{1}](111)$, $[\bar{1}2\bar{1}](111)$ and $[11\bar{2}](111)$  [see Fig.~\ref{fig:pure_ni3al}(a)],
due to the symmetry lowering as compared to Ni.
Slipping the system along these directions by $\boldsymbol{b}/2$, $\boldsymbol{b}/6$ and $\boldsymbol{b}/3$ burgers vectors
results in the formation of the APB, CSF and SISF, respectively.
The corresponding calculated GSFEs are given in Table~\ref{tab:pure_GSFEs}
and further shown in Figs.~\ref{fig:pure_ni3al}(b)-(d).
It can be seen that the $\gamma_{usf}$ for forming CSF (257 $mJ/m^2$) is much lower than $\gamma_{usf}$ for forming APB (819 $mJ/m^2$).
These results support the decomposition reaction of a $1/2 \langle 1\bar{1}0 \rangle$ dislocation into two Shockley partials on the (111) plane of Ni$_{3}$Al,
which can be described by
$1/2 \langle 01\bar{1} \rangle \rightarrow 1/6 \langle \bar{1}2\bar{1} \rangle +
1/6 \langle 11\bar{2} \rangle$~\cite{PhysRevB.101.024102} [see red arrows in Fig.~\ref{fig:pure_ni3al}(a)].
Thus, the $[\bar{1}2\bar{1}]$ direction exhibits the best ductility.
Moreover, the SISF exhibits the smallest $\gamma_{sf}$ of 68 $mJ/m^2$,
much smaller than those of APB (244 $mJ/m^2$) and CSF (214 $mJ/m^2$).
However, its formation needs to overcome the largest energy barrier ($\gamma_{usf}$=1339 $mJ/m^2$),
where the $\gamma_{usf}$ decreases in a sequence of
$[11\bar{2}](111)$  $>$  $[01\bar{1}](111)$  $>$  $[\bar{1}2\bar{1}](111)$.
This suggests that once SISF is produced, it is fairly kinetically stable and its slipping
becomes much more difficult. This might be connected with
the anomalous temperature dependence behaviour of the
YSA~\cite{RAE20071067, RAE2001125, KOVARIK2009839, VISWANATHAN20053041, MRYASOV20024545}.
Our predictions are in reasonable agreement with other DFT calculations and experimental estimations (see Table~\ref{tab:pure_GSFEs}).
The noticeable exception is that the experimentally estimated $\gamma_{sf}$ of SISF
from Karnthaler \emph{et al.}~\cite{KARNTHALER1996547} using weak-beam TEM images is somewhat too small (6$\pm$0.5 $mJ/m^2$).
However, the more recent experimental estimation by Knowles and Chen~\cite{KNOWLES200388} using the same technique
obtained a much larger value (35 $mJ/m^2$) that is in better agreement with the DFT calculated results.

\begin{figure}
\begin{center}
\includegraphics[width=0.47\textwidth]{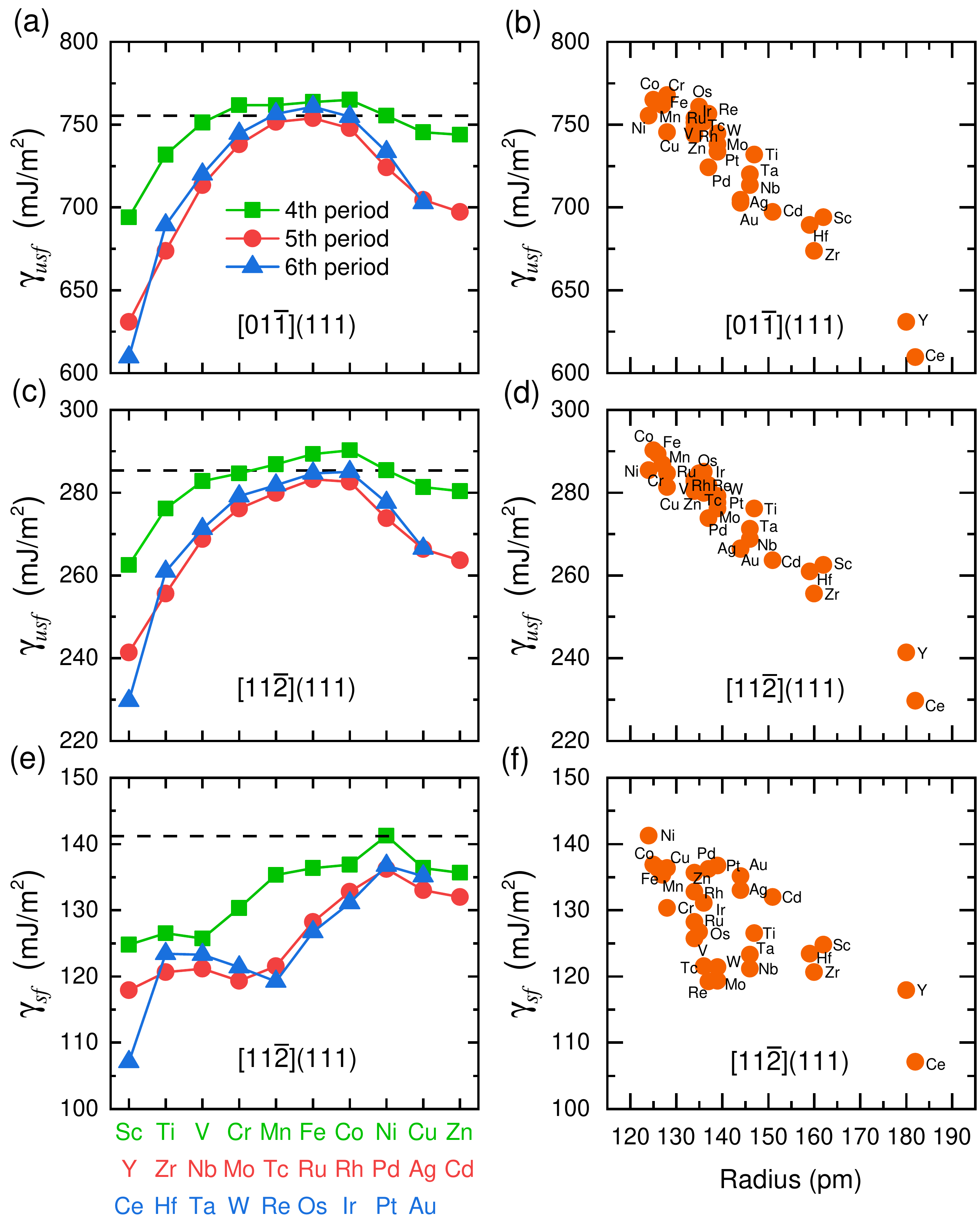}
\end{center}
\caption{The effects of considered alloying elements on the GSFEs of Ni
for the slip systems of
(a) $\gamma_{usf}$ of $[01\bar{1}](111)$
(c) $\gamma_{usf}$ of $[11\bar{2}](111)$
and (e) $\gamma_{sf}$ of $[11\bar{2}](111)$.
(b), (d) and (f) show the corresponding GSFEs as a function of
the radii of alloying elements.
The dashed line represents the $\gamma_{usf}$ and $\gamma_{sf}$ of pure Ni.
}
\label{fig:doped_ni}
\end{figure}

\subsection{Effects of alloying elements on the $\gamma_{usf}$ and $\gamma_{sf}$ of Ni}

To explore the effects of the alloying elements on the $\gamma_{usf}$ and $\gamma_{sf}$ of different slip systems in Ni,
twenty-nine alloying elements including most of transition-metal elements and Ce were considered.
Figure~\ref{fig:doped_ni} shows the calculated $\gamma_{usf}$ and $\gamma_{sf}$ with the addition of an alloying element on the SF plane.
The detailed values are summarized in Table~\ref{tab:dpoed_GSFEs}.
First, it is interesting to see that the calculated $\gamma_{usf}$ for
the $[01\bar{1}](111)$ and $[11\bar{2}](111)$ slip systems exhibit similar trends
as the atomic number of alloying elements increases along the period [Figs.~\ref{fig:doped_ni}(a) and (c)].
Specifically, as the element moves from the left to the right along the period,
the $\gamma_{usf}$ first increases,
then reaches the maximum around group VIII elements,
and eventually decrease as the atomic number increases further.
Second, as compared to Ni, most of alloying elements tend to reduce $\gamma_{usf}$
except for Cr, Mn, Fe, Co, Re and Os [see Figs.~\ref{fig:doped_ni}(a) and (c)].
Moreover, the alloying elements in the 5th and 6th periods are found to
exhibit more pronounced effects in decreasing $\gamma_{usf}$ than those in the 4th period.
In particular, Ce shows the strongest reduction of $\gamma_{usf}$.
This indicates that the addition of alloying elements in Ni render the deformation
of $[01\bar{1}](111)$ and $[11\bar{2}](111)$ slip systems more easily.

Moving to the $\gamma_{sf}$ for the $[11\bar{2}](111)$ slip system,
one can observe from Fig.~\ref{fig:doped_ni}(e)
that all alloying elements decrease the $\gamma_{sf}$,
with the effect being more pronounced as the alloying elements are far from Ni in the periodic table.
Similar to the effects of alloying elements on $\gamma_{usf}$,
the alloying elements in 5th and 6th periods also decrease $\gamma_{sf}$
more remarkably than the 4th period elements.
Although the elements in groups VIB-VIIB and the first two columns of group VIII
almost have a negligible effect (just $\pm$2\% reduction) on the $\gamma_{usf}$ along the $[01\bar{1}]$ and $[11\bar{2}]$ directions,
they have a considerable effect (about $\pm$15\% reduction) on the $\gamma_{sf}$ of $[11\bar{2}](111)$ slip system.
It is interesting to observe that as compared to the elements in the same period,
Mo and Re exhibit unusual reduction of $\gamma_{sf}$
due to their $d$-$d$ orbitals hybridizations with Ni~\cite{YU20095914}.
These results are consistent with the findings of Shang \emph{et al.}~\cite{Shang_2012}.
Since a small value of SF results in a low steady-state creep rate~\cite{MOHAMED1974779, ARGON1981293, YU20095914},
one can thus expect that the addition of Ce, Y, Re and Mo in Ni can improve creep resistance
and exhibit good solid solution strengthening effect.

Figures~\ref{fig:doped_ni}(b), (d) and (f) show
the calculated $\gamma_{usf}$ and $\gamma_{sf}$ of different slip systems
as a function of the atomic radii of alloying elements.
It is evident that the $\gamma_{usf}$ follows a linear behavior with respect to atomic radii
and the alloying elements with larger atomic radii in general exhibit more pronounced
reduction effect on $\gamma_{usf}$.
Among the considered elements, the effects of Y and Ce are strongest,
due to their large atomic radii difference with respect to that of Ni.
However, for $\gamma_{sf}$ the degree of linear correlation
with atomic radii is relatively decreased.
These results suggest that the strain effects induced by alloying
elements play a dominant role in reducing the $\gamma_{usf}$ and $\gamma_{sf}$ of Ni,
which are responsible for the trends manifested in Figs.~\ref{fig:doped_ni}(a), (c) and (e).

One note is in place here before closing this section.
We recall that with our employed supercell model the planar solute concentration is 6.25 at.\%.
Therefore, strictly speaking, our results and discussions apply to the situation of dilute solute concentration.
Exploring the effects of solute concentration on the GSFEs is, however, computationally too demanding
for all considered 29 alloying elements due to the rapid increase of configuration space, and is beyond the scope of the present work.
We shall also stress that in our work we focus only on the general trends of the effects of different alloying elements on the GSFEs of Ni and Ni$_3$Al.
Despite of this, it would still be interesting to see to what extent the trends obtained from our calculations
in dilute solute concentration apply to higher solute concentrations.
To this end, we compared our results in Fig.~\ref{figS1} of the Appendix to the available literature data for Al, Ti, Cr and Co
that were obtained using a relatively small supercell model~\cite{YU20095914,Shang_2012,DODARAN2021110326}.
Since the variations of stacking fault energies were found to almost follow a linear behavior with the solute concentration,
the overall trends across different alloying elements remain unchanged with the modification of solute concentrations,
i.e., in a sequence of Ti $>$ Cr $>$ Al $>$ Co for reducing the stacking fault energy (see Fig.~\ref{figS1}).
In this regard, we can say that the results obtained from our work still have a certain extrapolation capability to the not high solute concentration regime.
However, whether this still holds for other alloying elements remains to be carefully examined.

\subsection{Effects of alloying elements on the $\gamma_{usf}$ and $\gamma_{sf}$ of Ni$_3$Al}
	
\begin{figure}
\begin{center}
\includegraphics[width=0.4\textwidth]{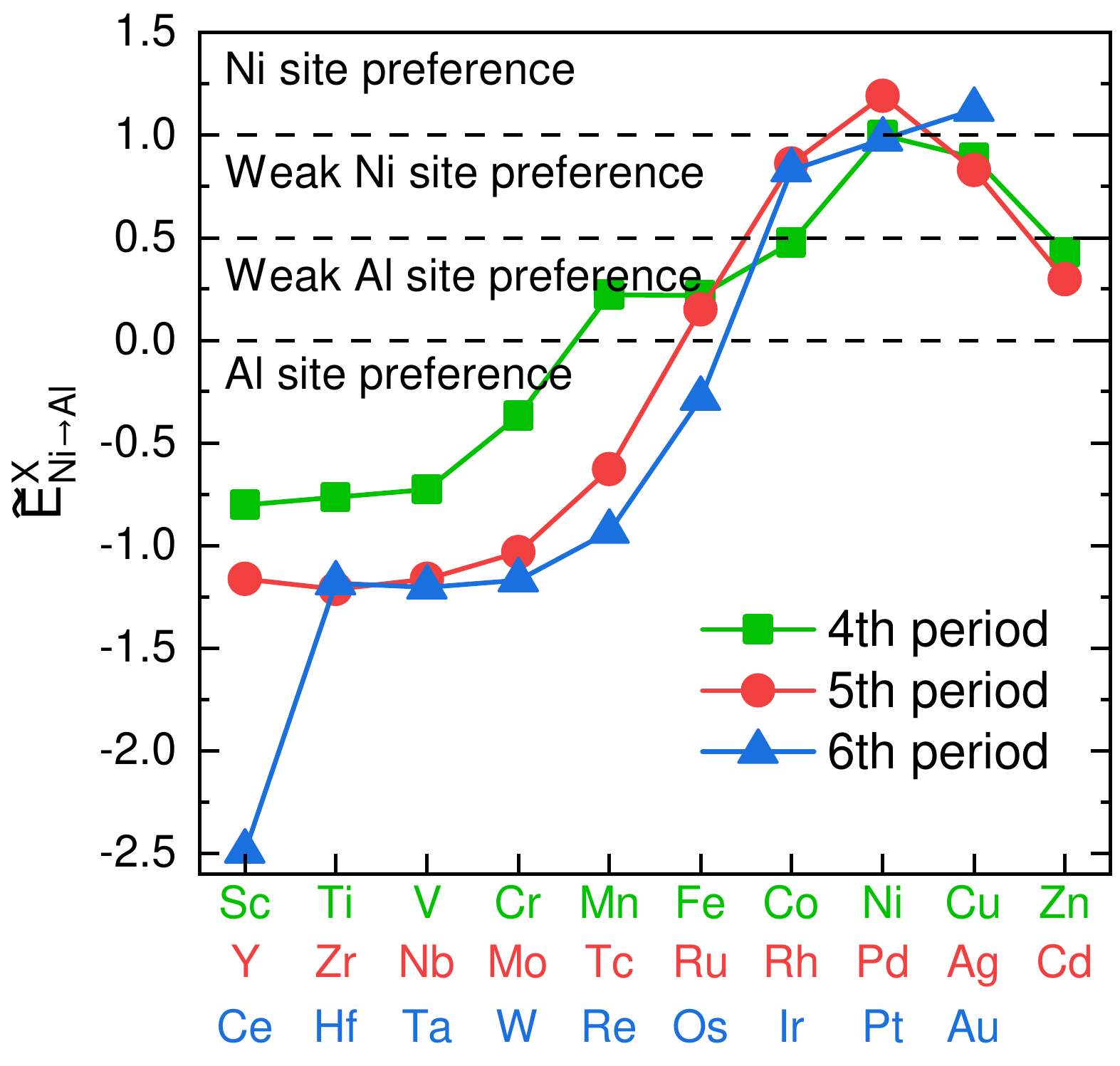}
\end{center}
\caption{Calculated normalized transfer energies of 29 alloying elements.
}
\label{fig:site_preference}
\end{figure}
	
\begin{figure}
\begin{center}
\includegraphics[width=0.49\textwidth]{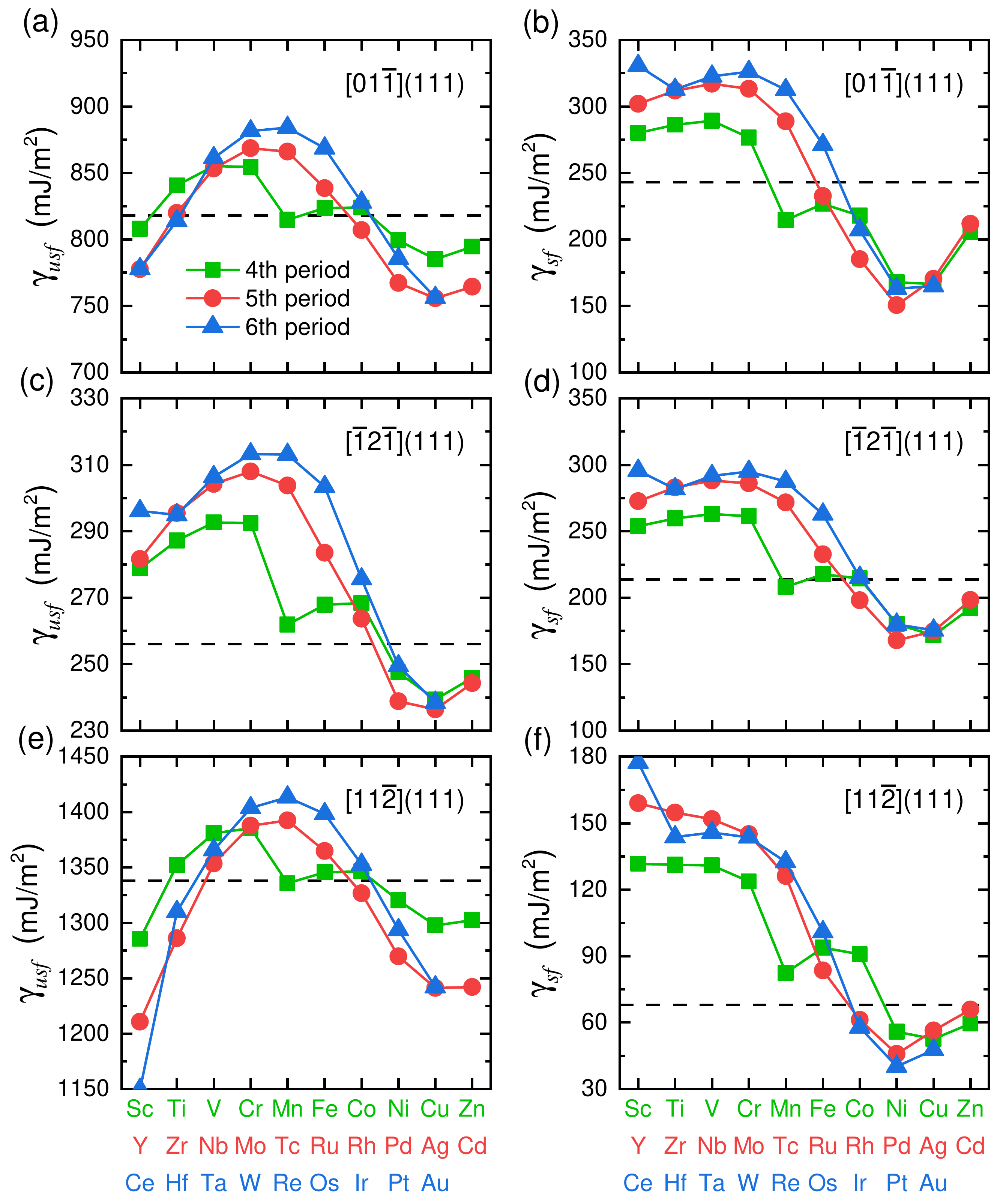}
\end{center}
\caption{The effects of considered alloying elements on the $\gamma_{usf}$ (left panels)  and  $\gamma_{sf}$ (right panels)of Ni$_3$Al
for the slip systems of
$[01\bar{1}](111)$ [(a) and (b)],
$[\bar{1}2\bar{1}](111)$ [(c) and (d)],
and $[11\bar{2}](111)$ [(e) and (f)].
The dashed line represents the $\gamma_{usf}$ and $\gamma_{sf}$ of pure Ni$_{3}$Al.
}
\label{fig:doped_ni3al}
\end{figure}

\begin{figure}
\begin{center}
\includegraphics[width=0.47\textwidth]{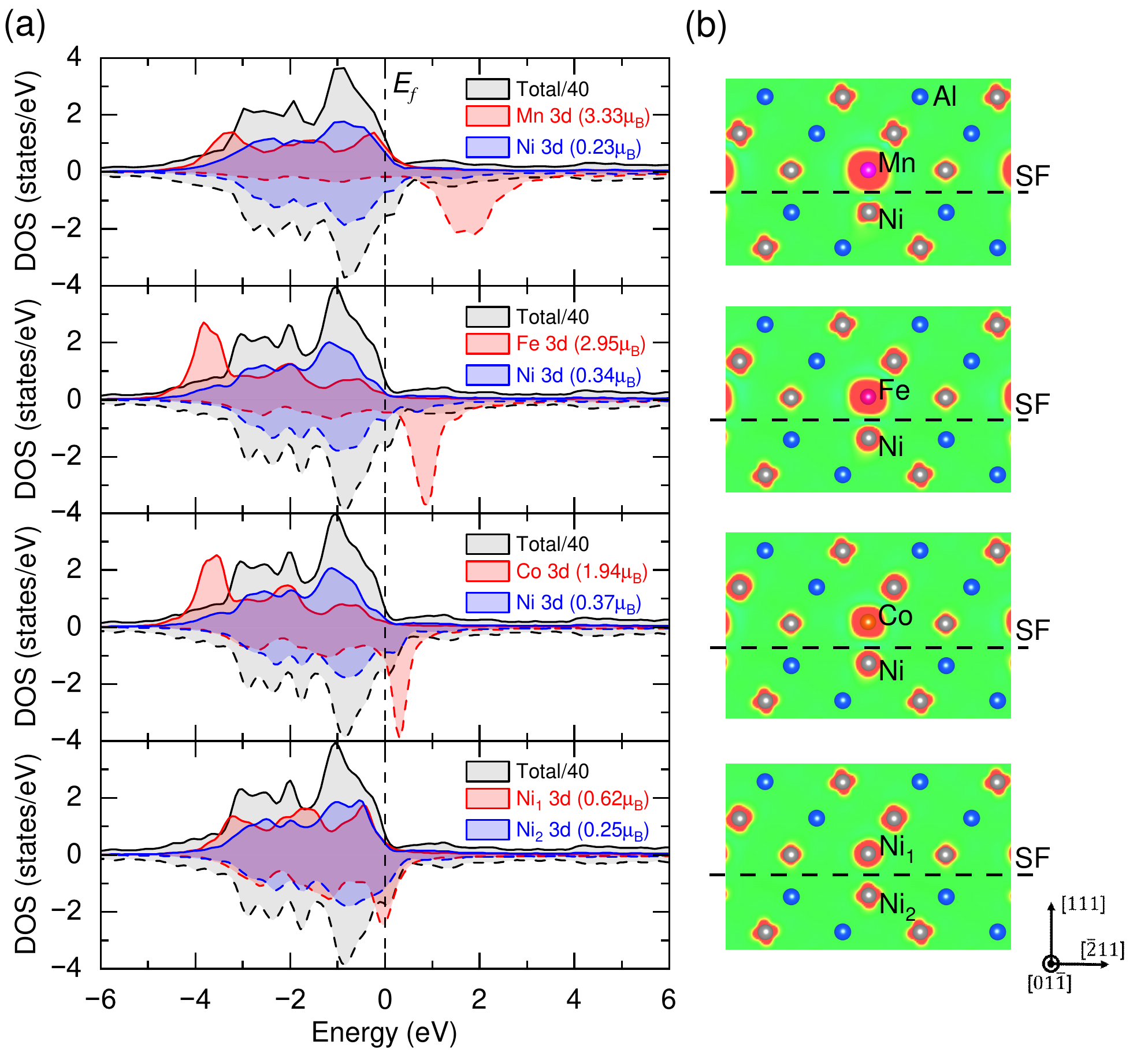}
\end{center}
\caption{(a) Spin-polarized local density of states (LDOSs) of alloying elements (Mn, Fe, Co and Ni)
and their first-nearest neighboring Ni atom
for the configuration used to calculate $\gamma_{usf}$
of $[11\bar{2}](111)$ slip in Ni$_{3}$Al.
The atomic positions are indicated in (b).
The Fermi energy has been aligned to zero.
For a better presentation, the total DOSs have been divided by a factor of 40.
The values in parentheses represent the calculated magnetic moments.
(b) The corresponding spin density isosurface (with an isovalue of 0.01 e/$\AA^3$)
for the ($01\bar{1}$) plane including the alloying elements.
}
\label{fig:112_dos}
\end{figure}

When an alloying element is added to ordered Ni$_{3}$Al phase,
it can either substitute the Ni site or the Al site.
Hence, one needs to first determine the site preference of alloying elements
before discussing their effects on the $\gamma_{usf}$ and $\gamma_{sf}$.
To this end, we employed the Wagner-Schottky model \cite{PhysRevB.56.3032, JIANG2006433}.
In this model, the site preference can be determined by the normalized transfer energy
\begin{align}\label{eq:e_1}
\widetilde{E}^{X}_{\rm Ni \rightarrow Al}=E^{X}_{\rm Ni \rightarrow Al}/E_{anti},
\end{align}
where $E_{anti}$ is the sum of the formation energy of an Al antisite defect and that of a Ni antisite defect.
$E^{X}_{\rm Ni \rightarrow Al}$ is the transfer energy that an alloying element ($X$) transfers from a Ni site to an Al site,
and at the same time, the Al atom goes to the Ni site where the $X$ atom initially occupied.
The transfer energy is defined as
\begin{align}\label{eq:e_2}
E^{X}_{\rm Ni \rightarrow Al} = E^{X}({\rm Al}) + E^{\rm Al}({\rm Ni}) - E^{X}({\rm Ni})-E_{\rm Ni_{3}Al},
\end{align}
where $E^{X}({\rm Al})$ and $E^{X}({\rm Ni})$ are the total energies of Ni$_{3}$Al with a $X$ atom at Al and Ni sites, respectively,
$E^{\rm Al}({\rm Ni})$ is the total energy of Ni$_{3}$Al with an Al antisite,
and $E_{\rm Ni_{3}Al}$ is the total energy of pure Ni$_{3}$Al.
According to the Wagner-Schottky model,
one can obtain that,
if $\widetilde{E}^{X}_{\rm Ni \rightarrow Al} < 0$,
the solute $X$ has a strong tendency to occupy the Al site,
if $\widetilde{E}^{X}_{\rm Ni \rightarrow Al} > 1.0$,
the solute has a strong tendency to occupy the Ni site,
if $0 < \widetilde{E}^{X}_{\rm Ni \rightarrow Al} < 0.5$,
the solute has a weak Al site preference,
and if $0.5 < \widetilde{E}^{X}_{\rm Ni \rightarrow Al} < 1.0$,
the solute has a weak Ni site preference.
	
Figure~\ref{fig:site_preference} shows the calculated normalized transfer energies of 29 alloying elements.
It is evident that Sc, Ti, V, Cr, Y, Zr, Nb, Mo, Tc, Ce, Hf, Ta, W, Re, and Os show a strong Al site preference,
while Pd, Pt, and Au exhibit a strong Ni site preference.
Mn, Fe, Co, Zn, Ru, and Cd display a weak Al site preference,
whereas Cu, Rh, Ag, and Ir show a weak Ni site preference.
Our results are consistent with previous studies based on the Wagner-Schottky
model~\cite{JIANG2006433, PhysRevB.51.4062, PhysRevB.55.856, WU2012436, LIU2017157}
and also agree well with the
findings of Chen \emph{et al.}~\cite{chen_modeling_2022} using a grand canonical dilute-solution model thermodynamic formalism.
In addition, the fact that Ti, Cr, Nb, Ta, W, and Re favor to occupy the Al site in Ni$_{3}$Al
agrees well with previous fist-principles
calculations based Monte-Carlo simulations~\cite{ZHU202354, PhysRevB.94.014116}
as well as the widely accepted experimental recognition~\cite{2006_Reed, doi10.1063/1.2956398}.
We note that the elements with $0 < \widetilde{E}^{X}_{\rm Ni \rightarrow Al} < 1$
exhibit a strong composition-dependent site preference.
For instance, they prefer to occupy the Ni site in Al-rich Ni$_{3}$Al,
but favor to occupy the Al site in Ni-rich Ni$_{3}$Al,
while in stoichiometric Ni$_{3}$Al they show no site preference such
that they randomly occupy the Al and Ni sites~\cite{EURICH201587, chen_modeling_2022}.

Considering that most of alloying elements tend to occupy the Al site in Ni$_{3}$Al,
in the following we focus on the effects of the alloying element occupying the Al site
on the $\gamma_{usf}$ and $\gamma_{sf}$ for three inequivalent slip systems.
The results are presented in Fig.~\ref{fig:doped_ni3al} and summarized in Table~\ref{tab:dpoed_GSFEs}.
It is interesting to find that the effects of alloying elements on $\gamma_{usf}$ or $\gamma_{sf}$
follow a similar trend for three slip systems.
Taking $\gamma_{usf}$ for example,
when the element moves from left to right in the same period,
the $\gamma_{usf}$ in general first increases, then reaches the maximum, and finally goes down.
For the elements in groups VB-VIIB and the first column of group VIII,
they have a striking impact on $\gamma_{usf}$ (except for Mn and Fe) [see Figs.~\ref{fig:doped_ni3al}(a), (c) and (e)].
Especially, the elements W and Re in the 6th period and Mo in the 5th period significantly increase the values of $\gamma_{usf}$.
Since large $\gamma_{usf}$ values make dislocation emits and plastic deformation more difficult~\cite{YU20095914},
one can expect that the addition of these elements to Ni$_3$Al can hinder the slip deformation.
On the other hand, the elements in groups IB-IIIB and the last column of group VIII have a negative effect on $\gamma_{usf}$,
in particular the elements Cu, Ag, and Au in group IB,
indicating that these elements can improve the slip deformation of Ni$_3$Al.

\begin{figure*}
	\begin{center}
		\includegraphics[width=0.8\textwidth]{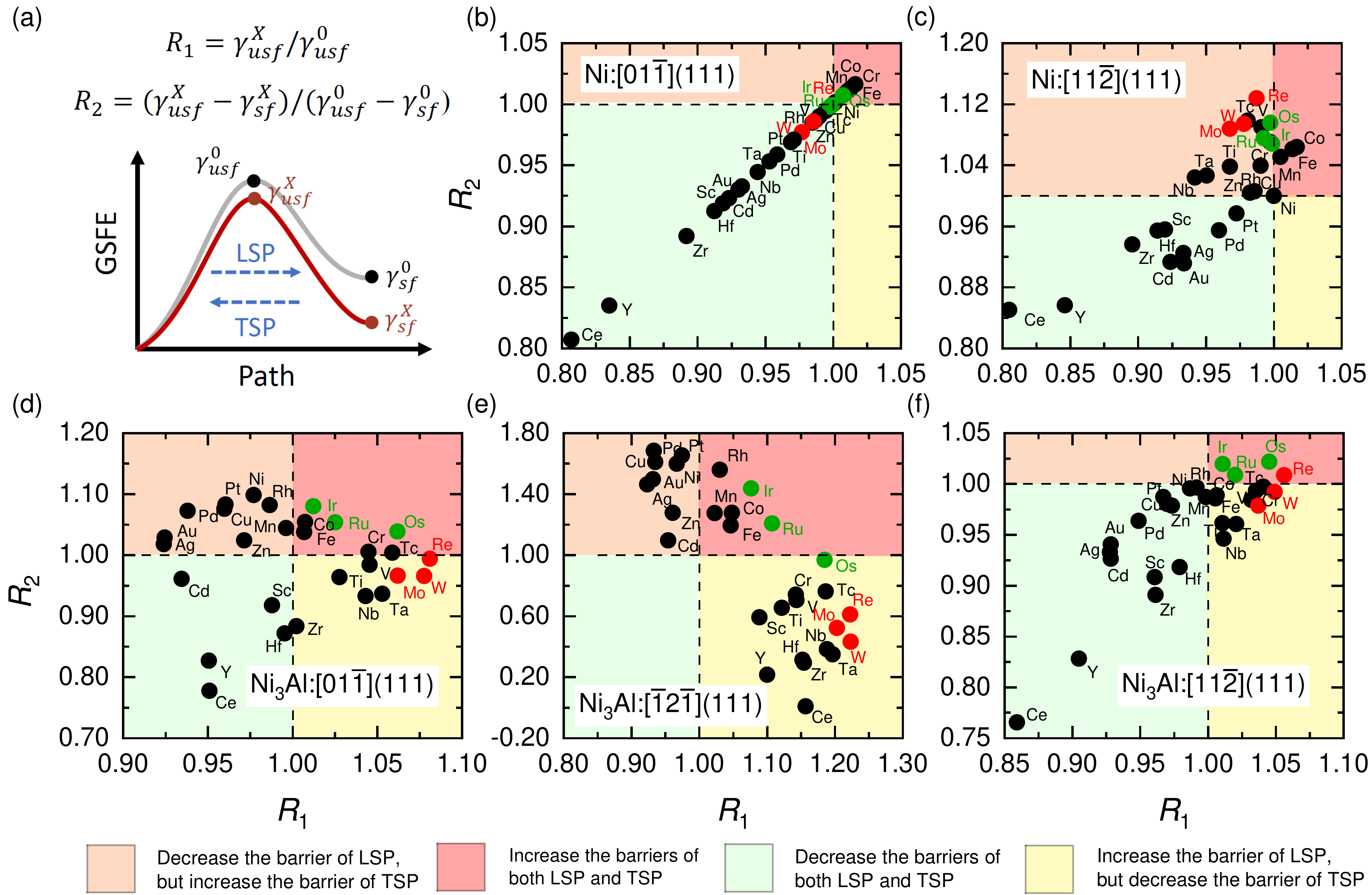}
	\end{center}
	\caption{(a) Sketch of the GSFE curve along a slip path with the LSP and TSP being indicated.
		(b)-(f) A summary of the effects of alloying elements on the barriers
		of LSP and TSP for different slip systems in Ni and Ni$_3$Al.
		(b) Ni: $[01\bar{1}](111)$,
		(c) Ni: $[11\bar{2}](111)$,
		(d) Ni$_{3}$Al: $[01\bar{1}](111)$,
		(e) Ni$_{3}$Al: $[\bar{1}2\bar{1}](111)$,
		(f) Ni$_{3}$Al: $[11\bar{2}](111)$.
	}
	\label{fig:slip_barrier}
\end{figure*}

For $\gamma_{sf}$,
the elements in groups IIIB-VIIB (except for Mn) increase $\gamma_{sf}$, especially the elements Ce, W and Re.
This means that these elements reduce the stability of
the APB configuration in the $[01\bar{1}](111)$ slip system,
the CSF configuration in the $[\bar{1}2\bar{1}](111)$ slip system,
and the SISF configuration in the $[11\bar{2}](111)$ slip system.
Our results are consistent with the findings of Yu \emph{et al.}~\cite{YU20095914, XIA2022104183}.
Since small values of SF lower the steady-state creep rate~\cite{MOHAMED1974779, ARGON1981293, YU20095914},
the elements in groups IIIB-VIIB would thus reduce the creep strength of Ni$_{3}$Al.

By examining calculated $\gamma_{usf}$ and $\gamma_{sf}$ of Ni$_3$Al versus atom radii (not shown),
we found that they do not simply follow a linear behavior as the case in Ni.
Thus, besides the strain effects,
electronic structures changed by the alloying elements also take effects in
determining the values of $\gamma_{usf}$ and $\gamma_{sf}$ for Ni$_3$Al.
Furthermore, one notices that among the considered elements,
Mn, Fe and Co exhibit abnormal behaviors (see Fig.~\ref{fig:doped_ni3al}).
Our electronic structure analysis demonstrates that these abnormal behaviors
originate from their strong spin polarizations.
Just taking $\gamma_{usf}$ of $[11\bar{2}](111)$ slip in Ni$_{3}$Al as an example,
we plot in Fig.~\ref{fig:112_dos} the LDOSs of alloying elements (Mn, Fe, Co and Ni)
and their first-nearest neighboring Ni atom as well as corresponding spin density isosurfaces.
It is evident that the spin polarization decreases in the order of Mn, Fe, Co and Ni,
yielding local magnetic moments of
3.33 $\mu_\text{B}$, 2.95 $\mu_\text{B}$, 1.94 $\mu_\text{B}$ and 0.62 $\mu_\text{B}$, respectively.
By contrast, the other alloying elements do not exhibit magnetic couplings with the neighboring Ni atoms.

\subsection{Effects of alloying elements on normal slip barriers}

The slip deformation is a reaction process with an energy barrier ($\gamma_{usf}$) to overcome to form stable SFs,
whose stability is described by $\gamma_{sf}$.
In order to better describe the effects of alloying elements on the SFs,
here, we defined the process from a perfect supercell slip to the formation of a stable SF configuration as the leading slip process (LSP)
and its inverse process was defined as the tailing slip process (TSP) [see Fig.~\ref{fig:slip_barrier}(a)].
We notice that simply comparing $\gamma_{sf}$ and $ \gamma_{usf}$
is not sufficient to capture the overall effects of different alloying elements on GSFEs in an accurate manner.
To address this issue, we have proposed renormalized
$R_{1}=\gamma^{X}_{usf}/\gamma^{0}_{usf}$
and $R_{2}=(\gamma^{X}_{usf}-\gamma^{X}_{sf})/(\gamma^{0}_{usf}-\gamma^{0}_{sf})$ indices
to effectively characterize the effects of alloying elements on the slip barriers of the LSP and TSP, respectively.
Here, $\gamma^{X}_{usf}$ and $\gamma^{0}_{usf}$ represent the unstable stacking fault energies with and without alloying element $X$, respectively,
whereas $\gamma^{X}_{sf}$ and $\gamma^{0}_{sf}$ denote the stable stacking fault energies with and without alloying element $X$, respectively.
According to this definition, one can obtain that
the indices $R_{1}$ ($R_{2}$) greater than 1
mean that the alloying elements increase the barrier of the LSP (TSP),
which makes deformation more difficult.
On the contrary, when the indices are less than 1, the deformation becomes easier with the addition of alloying elements.
In this way, the different alloying elements with similar effects can well be classified into the same $R_{1}$-$R_{2}$ quadrant.

The clarification results are compiled in Fig.~\ref{fig:slip_barrier} in terms of $R_{1}$ and $R_{2}$.
From the plot it is evident that among all considered alloying elements
Re is an excellent strengthening element that significantly increases
the barrier of the TSP of the $[01\bar{1}](111)$ slip system for the Ni phase,
and also enhances the barrier of the LSP of the $[01\bar{1}](111)$, $[\bar{1}2\bar{1}](111)$, $[11\bar{2}](111)$ three slip systems for the Ni$_{3}$Al phase.
However, it shows a negligible or even slight negative impact
on the LSP of $[01\bar{1}](111)$ slip system in Ni
and the TSP of $[\bar{1}2\bar{1}](111)$ slip system in Ni$_{3}$Al.
We note that W and Mo exhibit similar effects as Re.
The element of Os is almost distributed in the red area (see Fig.~\ref{fig:slip_barrier}),
which suggests that it shows strengthening effects on both LSP and TSP for Ni and Ni$_3$Al.
Our findings are consistent with the design strategy recently proposed by
Bensheng \emph{et al.}~\cite{WEI2022118336} who
suggested to replace Re by Os in order to improve creep resistance and
phase stability of nickel-based single crystal superalloys.
It turns out that the newly designed Os-containing superalloy indeed shows a lower creep rate
than commercial Re-containing CMSX-4 alloy at 980 $^{\circ}$C/200 MPa/100 hours~\cite{WEI2022118336}.
In addition, we found that Ru and Ir exhibit similar effects as Os.
This might explain why the Ru element has been included in the fourth-generation
(EPM-102~\cite{10.7449/2004/Superalloys_2004_15_24} and TMS-138~\cite{ZHANG20035073})
and fifth-generation NSCSs (TMS-173~\cite{Kobayashi2005CreepSO}),
and the Ir element has been included in the sixth-generation NSCSs (TMS-238~\cite{10.1007/978-3-030-51834-9_12, 10.1002/9781118516430.ch21}).
By contrast, the rare-earth elements Y and Ce dramatically decrease $R_{1}$ and $R_{2}$,
indicating that they are not good alloying elements
for improving the slip barrier of Ni or Ni$_{3}$Al.
Despite of this, a certain amount of Y or Ce has been added to superalloys
because of their strong deoxidizing and desulfurizing abilities~\cite{CAO2021260}.

\section{Conclusions}

In summary, we have systematically studied the effects of
29 alloying elements on the GSFEs of different slip systems in Ni and Ni$_{3}$Al
through a comprehensive first-principles calculations based on the alias shear method.
The conclusions drawn from this study are as follows:

($i$) For Ni, except for magnetic elements Mn, Fe and Co,
most of alloying elements decrease $\gamma_{usf}$ of
 $[01\bar{1}](111)$ and $[11\bar{2}](111)$ slip systems
and also decrease $\gamma_{sf}$ of the $[11\bar{2}](111)$ slip system.
The reduction effects show a strong correlation with the inverse of atom radii.

($ii$) For Ni$_{3}$Al, most of alloying elements in groups IIIB-VIIB show a strong Al site preference.
Except for Mn and Fe, the elements in groups VB-VIIB and the first column of group VIII increase the value of
$\gamma_{usf}$ of different slip systems of the Ni$_{3}$Al phase,
which makes the slip deformation and dislocation emits difficult.
However, the elements in groups IIIB-VIIB also increase the value of $\gamma_{sf}$,
and thus reduce the stability of APB, CSF and SISF configuration of the Ni$_{3}$Al phase.

($iii$) The alloying elements have been suitably clarified into four quadrants in terms of
the two proposed indices $R_1$ and $R_2$ (see Fig.~\ref{fig:slip_barrier}).
We obtained that Re is an excellent strengthening alloying element that significantly
increases the slip barrier of the tailing slip process for the Ni phase,
and also enhances the slip barrier of the leading slip process of three slip systems for the Ni$_{3}$Al phase.
W and Mo exhibit similar effects as Re.

($iv$) We predicted that Os, Ru, and Ir are also good strengthening alloying elements,
which show the strengthening effects on both the leading and tailing slip processes for Ni and Ni$_{3}$Al.

We anticipate that our established dictionary of the effects of various alloying elements on the GSFEs of both Ni and Ni$_3$Al phases
and new findings would be appreciated by the broad community for guiding the design of the next-generation high-performance Ni-based single crystal superalloys.

\vspace{10mm}
	
\setlength{\tabcolsep}{6.5pt}
\setlength{\LTcapwidth}{\textwidth}
\renewcommand\arraystretch{1.2}
\begin{longtable*}{ccllllllll}
\caption{A summary of calculated stable stacking fault energies ($\gamma_{sf}$) and unstable stacking fault energies ($\gamma_{usf}$)
of different slip systems in Ni and Ni$_{3}$Al with the addition of alloying elements.
The available literature data are also given.}
\label{tab:dpoed_GSFEs} \\
		
\hline \hline
	\multirow{2}{*}{Elements} & Ni: $[01\bar{1}](111)$ & \multicolumn{2}{l}{Ni: $[11\bar{2}](111)$}   & \multicolumn{2}{l}{Ni$_{3}$Al: $[01\bar{1}](111)$}   & \multicolumn{2}{l}{Ni$_{3}$Al: $[\bar{1}2\bar{1}](111)$}   & \multicolumn{2}{l}{Ni$_{3}$Al: $[11\bar{2}](111)$}   \\ \cline{2-10}
		& $\gamma_{usf}$   & $\gamma_{usf}$   & $\gamma_{sf}$   & $\gamma_{usf}$   & $\gamma_{sf} (\gamma_{\rm APB})$   & $\gamma_{usf}$   & $\gamma_{sf} (\gamma_{\rm CSF})$   & $\gamma_{usf}$       & $\gamma_{sf} (\gamma_{\rm SISF})$   \\
		\hline
		\endfirsthead
		
		\multicolumn{10}{l}{{(Continued)}} \\
		\hline
    	\multirow{2}{*}{Elements} & Ni: $[01\bar{1}](111)$ & \multicolumn{2}{l}{Ni: $[11\bar{2}](111)$}   & \multicolumn{2}{l}{Ni$_{3}$Al: $[01\bar{1}](111)$}   & \multicolumn{2}{l}{Ni$_{3}$Al: $[\bar{1}2\bar{1}](111)$}   & \multicolumn{2}{l}{Ni$_{3}$Al: $[11\bar{2}](111)$}   \\ \cline{2-10}
		& $\gamma_{usf}$   & $\gamma_{usf}$   & $\gamma_{sf}$   & $\gamma_{usf}$   & $\gamma_{sf}(\gamma_{\rm APB})$   & $\gamma_{usf}$   & $\gamma_{sf}(\gamma_{\rm CSF})$   & $\gamma_{usf}$       & $\gamma_{sf}(\gamma_{\rm SISF})$   \\
		\hline
		\endhead
		
		
		\hline \hline
		\endlastfoot
		
Sc & 694  & 263  & 125 & 808 & 280 & 279 & 254 & 1285 & 132  \\
   &      &      & 108 \cite{Shang_2012} &     &      &      &      &     &          \\
Ti & 732  & 276  & 127 & 841 & 286 & 287 & 260 & 1352 & 131          \\
   &      &      & 113 \cite{Shang_2012} & 839\cite{YU201238} &     & 308\cite{YU201238} &     & 1438\cite{YU201238}  &        \\
V  & 751  & 283  & 126 & 855 & 289 & 293 & 263 & 1381 & 131                            \\
   &      &      & 113\cite{Shang_2012}  &     &      &       &     &      &         \\
Cr & 768  & 285  & 130  & 855 & 277 & 292 & 261 & 1385 & 124        \\
   &      & 248\cite{YANG2020109682} &\makecell[tl]{119 \cite{Shang_2012},\\110\cite{YANG2020109682},\\107\cite{YU20095914}} &   &    &    &    &     &      \\
Mn & 762  & 287  & 135  & 815 & 215 & 262 & 208 & 1336 & 82                             \\
   &      & 274\cite{doi10.1063/1.2051793}  &\makecell[tl]{124 \cite{Shang_2012},\\86\cite{doi10.1063/1.2051793}}  &   &   &   &   &      &          \\
Fe & 764  & 289  & 136  & 824 & 227 & 268 & 218 & 1345 & 94        \\
   &      & 280\cite{doi10.1063/1.2051793}  & \makecell[tl]{125 \cite{Shang_2012},\\86\cite{doi10.1063/1.2051793}}  &  &   &   &  &      &        \\
Co & 765  & 290  & 137  & 824 & 218 & 268 & 215 & 1347 & 91                             \\
   &      & 234\cite{YANG2020109682}      & \makecell[tl]{127 \cite{Shang_2012},\\110\cite{YANG2020109682},\\124\cite{YU20095914}} &      &      &      &     &     &      \\
Ni & 755  & 285  & 141  & 799  & 168  & 247  & 180  & 1320  & 56      \\
   &      &      & 132 \cite{Shang_2012}   &   &   &   &   &   &    \\
Cu & 745  & 281  & 136  & 785  & 167  & 239  & 172  & 1298  & 53      \\
   &      & 256\cite{doi10.1063/1.2051793}   & \makecell[tl]{126 \cite{Shang_2012},\\86\cite{doi10.1063/1.2051793}} &   &   &   &  &      &    \\
Zn & 744  & 280  & 136  & 795  & 206  & 246  & 192  & 1302  & 59      \\
   &      &      & 125 \cite{Shang_2012}     &      &       &     &     &     &    \\
Y  & 631  & 241  & 118  & 778  & 302  & 282  & 273  & 1211  & 159     \\
   & 589\cite{HU2020155799}    & 220\cite{HU2020155799}     & \makecell[tl]{96 \cite{Shang_2012},\\130\cite{HU2020155799}}   & 731\cite{HU2020155799}      & 302\cite{HU2020155799}     &     &     & 1083\cite{HU2020155799}  & 81\cite{HU2020155799}  \\
Zr & 674  & 256  & 121  & 820  & 312  & 295  & 283  & 1286  & 155   \\
   & 656\cite{HU2020155799}   & 244\cite{HU2020155799} & \makecell[tl]{101 \cite{Shang_2012},\\135\cite{HU2020155799}}  & 858\cite{HU2020155799}  & 365\cite{HU2020155799} &   &    & 1350\cite{HU2020155799}  & 316\cite{HU2020155799} \\
Nb & 713  & 269  & 121  & 853  & 317  & 304  & 288  & 1353  & 152    \\
   &      & 223\cite{doi10.1063/1.2051793}    & \makecell[tl]{104\cite{Shang_2012},\\59\cite{doi10.1063/1.2051793}}   &   &   &   &    &    &   \\
Mo & 738  & 276  & 119  & 869  & 313  & 308  & 286  & 1388  & 145    \\
   &      & 280\cite{XIA2022104183}   & \makecell[tl]{ 103\cite{Shang_2012},\\55\cite{XIA2022104183},\\100\cite{YU20095914}}  & 1034\cite{XIA2022104183}   & 531\cite{XIA2022104183} &   &    & 1553\cite{XIA2022104183}   & 238\cite{XIA2022104183} \\
Tc & 752  & 280  & 122  & 866  & 289  & 304  & 272  & 1393  & 126    \\
   &      &      & 105\cite{Shang_2012}    &   &   &   &    &   &   \\
Ru & 754  & 283  & 128  & 839  & 233  & 284  & 233  & 1365  & 83     \\
   &      & 308\cite{XIA2022104183}   & \makecell[tl]{116\cite{Shang_2012},\\95\cite{XIA2022104183},\\108\cite{YU20095914}}   & \makecell[tl]{ 905\cite{XIA2022104183},\\837\cite{YU201238}}   & 185\cite{XIA2022104183}  & \makecell[tl]{ 339\cite{XIA2022104183},\\294\cite{YU201238}}  &242\cite{XIA2022104183} & \makecell[tl]{ 1456\cite{XIA2022104183},\\1350\cite{YU201238}} & 29\cite{XIA2022104183}  \\
Rh & 748  & 283  & 133  & 807  & 185  & 264  & 198  & 1327  & 61   \\
   &      &      & 122\cite{Shang_2012}   &       &      &     &     &     &    \\
Pd & 724  & 274  & 136  & 767  & 151  & 239  & 168  & 1270  & 46   \\
   &      &      & 127\cite{Shang_2012}   &       &      &     &     &     &    \\
Ag & 705  & 266  & 133  & 756  & 170  & 236  & 175  & 1241  & 56   \\
   &      &      &      &      &      &      &      &       &      \\
Cd & 697  & 264  & 132  & 764  & 212  & 244  & 198  & 1242  & 66   \\
   &      &      &      &      &      &      &      &       &      \\
Ce & 610  & 230  & 107  & 778  & 331  & 296  & 296  & 1149  & 177  \\
   &      &      &      &      &      &      &      &       &      \\
Hf & 689  & 261  & 123  & 814  & 313  & 295  & 282  & 1310  & 144  \\
   & 689\cite{HU2020155799}   & 253\cite{HU2020155799}  & \makecell[tl]{ 106\cite{Shang_2012},\\140\cite{HU2020155799}} & 914\cite{HU2020155799} & 373\cite{HU2020155799} &   &  & 1404\cite{HU2020155799}  & 300\cite{HU2020155799}  \\
Ta & 720  & 271  & 123  & 861  & 323  & 306  & 292  & 1366  & 146  \\
   &      & 276\cite{XIA2022104183}    & \makecell[tl]{ 108\cite{Shang_2012},\\86\cite{XIA2022104183}} & \makecell[tl]{1013\cite{XIA2022104183},\\908\cite{YU201238}} & 585\cite{XIA2022104183} & \makecell[tl]{ 588\cite{XIA2022104183},\\337\cite{YU201238}} & 394\cite{XIA2022104183} & \makecell[tl]{ 1479\cite{XIA2022104183},\\1470\cite{YU201238}} & 347\cite{XIA2022104183} \\
W  & 745  & 279  & 121  & 881  & 326  & 313  & 295  & 1404  & 144  \\
   &      & \makecell[tl]{257\cite{doi10.1063/1.2051793},\\289\cite{XIA2022104183}}       & \makecell[tl]{ 105\cite{Shang_2012},\\38\cite{doi10.1063/1.2051793},\\59\cite{XIA2022104183},\\103\cite{YU20095914}}  & \makecell[tl]{1094\cite{XIA2022104183},\\939\cite{YU201238}} & 599\cite{XIA2022104183} & \makecell[tl]{611\cite{XIA2022104183},\\350\cite{YU201238}} & 425\cite{XIA2022104183} & \makecell[tl]{1636\cite{XIA2022104183},\\1531\cite{YU201238}} & 281\cite{XIA2022104183} \\
Re & 757  & 282  & 119  & 884  & 313  & 313  & 287  & 1413  & 132  \\
   &      & 301\cite{XIA2022104183}   & \makecell[tl]{103\cite{Shang_2012},\\48\cite{XIA2022104183},\\100\cite{YU20095914}}     & \makecell[tl]{1079\cite{XIA2022104183},\\941\cite{YU201238}} & 494\cite{XIA2022104183} & \makecell[tl]{524\cite{XIA2022104183},\\366\cite{YU201238}} & 432\cite{XIA2022104183} & \makecell[tl]{1628\cite{XIA2022104183},\\1533\cite{YU201238}} & 142\cite{XIA2022104183} \\
Os & 761  & 285  & 127  & 869  & 271  & 303  & 263  & 1398  & 101  \\
   &      &      & 113\cite{Shang_2012}    &    &   &    &   &   &    \\
Ir & 755  & 285  & 131  & 828  & 207  & 276  & 215  & 1353  & 58   \\
   &      &      & 120\cite{Shang_2012}    &    &   &    &   &   &    \\
Pt & 734  & 278  & 137  & 786  & 163  & 249  & 180  & 1294  & 40   \\
   &      &      & 128\cite{Shang_2012}    &    &   &    &   &   &    \\
Au & 703  & 266  & 135  & 756  & 165  & 238  & 176  & 1242  & 48   \\
\end{longtable*}
	
\begin{acknowledgments}
This work was supported by the National Key R\&D Program of China (No.~2021YFB3501503),
the National Science Fund for Distinguished Young Scholars (No. 51725103),
and the funding of National Science and Technology Major Project (J2019-VI-0004-0118, J2019-VI-0019-0134).
All calculations were performed on the high performance computational cluster at the Shenyang National University Science and Technology Park.
\end{acknowledgments}

\appendix*
\setcounter{figure}{0} \renewcommand{\thefigure}{\arabic{figure}}
\renewcommand{\thefigure}{A\arabic{figure}}
\setcounter{figure}{0}

\section{}

\begin{figure}[h!]
\begin{center}
\includegraphics[width=0.35\textwidth]{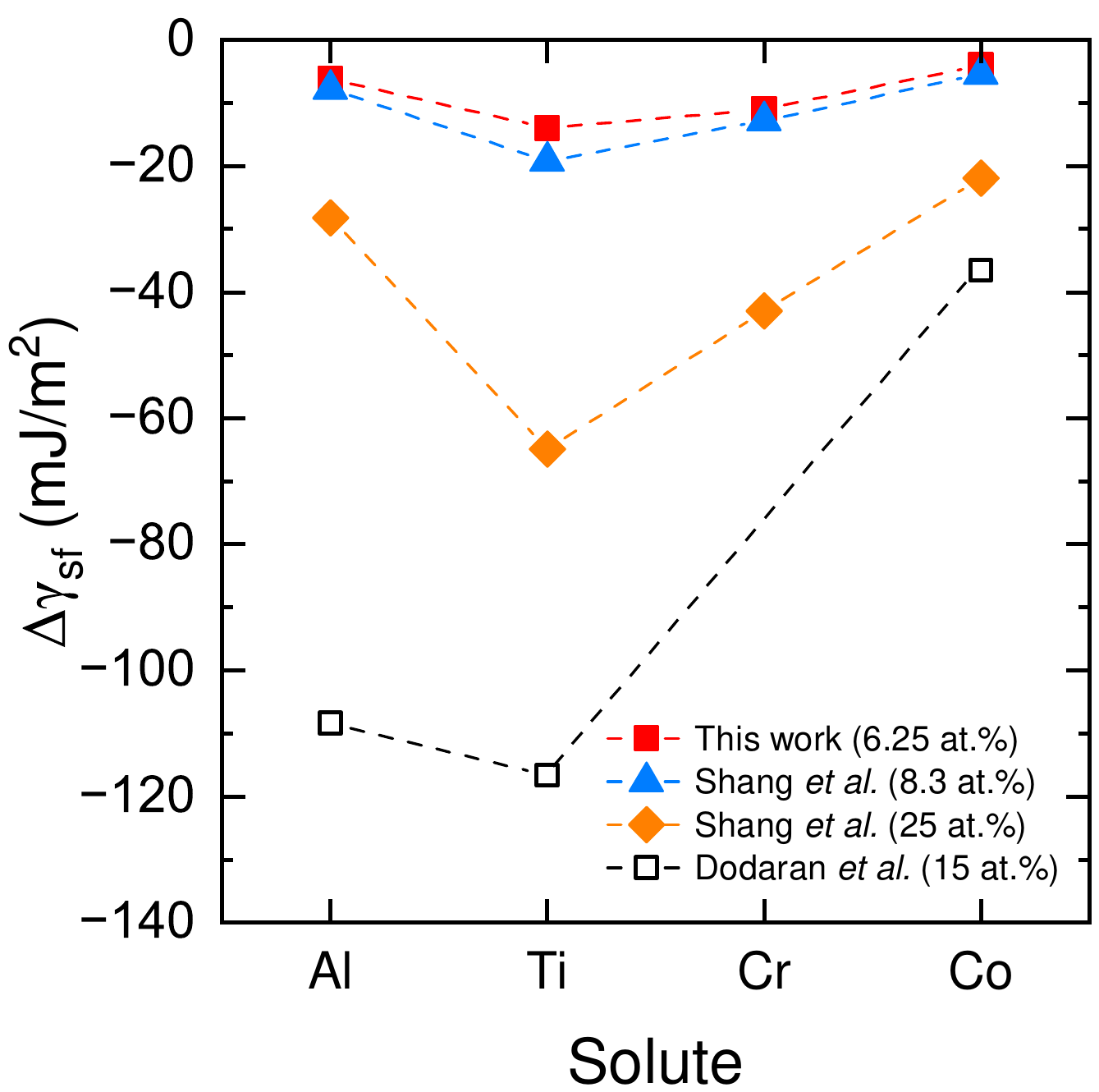}
\end{center}
\caption{Variation of stacking faults energies with solutes for four different planar solute concentrations (in parentheses).
The available literature data are taken from Refs.~\cite{YU20095914,Shang_2012, DODARAN2021110326}.
}
\label{figS1}
\end{figure}

\newpage
\bibliography{reference}

\providecommand{\noopsort}[1]{}\providecommand{\singleletter}[1]{#1}%
\begin{thebibliography}{81}%
\makeatletter
\providecommand \@ifxundefined [1]{%
 \@ifx{#1\undefined}
}%
\providecommand \@ifnum [1]{%
 \ifnum #1\expandafter \@firstoftwo
 \else \expandafter \@secondoftwo
 \fi
}%
\providecommand \@ifx [1]{%
 \ifx #1\expandafter \@firstoftwo
 \else \expandafter \@secondoftwo
 \fi
}%
\providecommand \natexlab [1]{#1}%
\providecommand \enquote  [1]{``#1''}%
\providecommand \bibnamefont  [1]{#1}%
\providecommand \bibfnamefont [1]{#1}%
\providecommand \citenamefont [1]{#1}%
\providecommand \href@noop [0]{\@secondoftwo}%
\providecommand \href [0]{\begingroup \@sanitize@url \@href}%
\providecommand \@href[1]{\@@startlink{#1}\@@href}%
\providecommand \@@href[1]{\endgroup#1\@@endlink}%
\providecommand \@sanitize@url [0]{\catcode `\\12\catcode `\$12\catcode
  `\&12\catcode `\#12\catcode `\^12\catcode `\_12\catcode `\%12\relax}%
\providecommand \@@startlink[1]{}%
\providecommand \@@endlink[0]{}%
\providecommand \url  [0]{\begingroup\@sanitize@url \@url }%
\providecommand \@url [1]{\endgroup\@href {#1}{\urlprefix }}%
\providecommand \urlprefix  [0]{URL }%
\providecommand \Eprint [0]{\href }%
\providecommand \doibase [0]{http://dx.doi.org/}%
\providecommand \selectlanguage [0]{\@gobble}%
\providecommand \bibinfo  [0]{\@secondoftwo}%
\providecommand \bibfield  [0]{\@secondoftwo}%
\providecommand \translation [1]{[#1]}%
\providecommand \BibitemOpen [0]{}%
\providecommand \bibitemStop [0]{}%
\providecommand \bibitemNoStop [0]{.\EOS\space}%
\providecommand \EOS [0]{\spacefactor3000\relax}%
\providecommand \BibitemShut  [1]{\csname bibitem#1\endcsname}%
\let\auto@bib@innerbib\@empty
\bibitem [{\citenamefont {Reed}(2006)}]{2006_Reed}%
  \BibitemOpen
  \bibfield  {author} {\bibinfo {author} {\bibfnamefont {R.~C.}\ \bibnamefont
  {Reed}},\ }\href {\doibase 10.1017/CBO9780511541285} {\emph {\bibinfo {title}
  {The Superalloys Fundamentals and Applications}}}\ (\bibinfo  {publisher}
  {Cambridge University Press},\ \bibinfo {year} {2006})\BibitemShut {NoStop}%
\bibitem [{\citenamefont {Royce}(2015)}]{2015_Royce}%
  \BibitemOpen
  \bibfield  {author} {\bibinfo {author} {\bibfnamefont {R.}~\bibnamefont
  {Royce}},\ }\href@noop {} {\emph {\bibinfo {title} {The jet engine}}}\
  (\bibinfo  {publisher} {John Wiley \& Sons},\ \bibinfo {year}
  {2015})\BibitemShut {NoStop}%
\bibitem [{\citenamefont {Pollock}\ and\ \citenamefont
  {Tin}(2006{\natexlab{a}})}]{2006_Pollock}%
  \BibitemOpen
  \bibfield  {author} {\bibinfo {author} {\bibfnamefont {T.}~\bibnamefont
  {Pollock}}\ and\ \bibinfo {author} {\bibfnamefont {S.}~\bibnamefont {Tin}},\
  }\href {\doibase 10.2514/1.18239} {\bibfield  {journal} {\bibinfo  {journal}
  {Journal of Propulsion and Power}\ }\textbf {\bibinfo {volume} {22}},\
  \bibinfo {pages} {361} (\bibinfo {year} {2006}{\natexlab{a}})}\BibitemShut
  {NoStop}%
\bibitem [{\citenamefont {Caron}\ and\ \citenamefont
  {Khan}(1999)}]{1999_Caron}%
  \BibitemOpen
  \bibfield  {author} {\bibinfo {author} {\bibfnamefont {P.}~\bibnamefont
  {Caron}}\ and\ \bibinfo {author} {\bibfnamefont {T.}~\bibnamefont {Khan}},\
  }\href {\doibase 10.1016/S1270-9638(99)00108-X} {\bibfield  {journal}
  {\bibinfo  {journal} {Aerospace Science and Technology}\ }\textbf {\bibinfo
  {volume} {3}},\ \bibinfo {pages} {513} (\bibinfo {year} {1999})}\BibitemShut
  {NoStop}%
\bibitem [{\citenamefont {Pollock}\ and\ \citenamefont
  {Tin}(2006{\natexlab{b}})}]{doi10.2514/1.18239}%
  \BibitemOpen
  \bibfield  {author} {\bibinfo {author} {\bibfnamefont {T.~M.}\ \bibnamefont
  {Pollock}}\ and\ \bibinfo {author} {\bibfnamefont {S.}~\bibnamefont {Tin}},\
  }\href {\doibase 10.2514/1.18239} {\bibfield  {journal} {\bibinfo  {journal}
  {Journal of Propulsion and Power}\ }\textbf {\bibinfo {volume} {22}},\
  \bibinfo {pages} {361} (\bibinfo {year} {2006}{\natexlab{b}})}\BibitemShut
  {NoStop}%
\bibitem [{\citenamefont {Karnthaler}\ \emph {et~al.}(1996)\citenamefont
  {Karnthaler}, \citenamefont {M\"{u}hlbacher},\ and\ \citenamefont
  {Rentenberger}}]{KARNTHALER1996547}%
  \BibitemOpen
  \bibfield  {author} {\bibinfo {author} {\bibfnamefont {H.}~\bibnamefont
  {Karnthaler}}, \bibinfo {author} {\bibfnamefont {E.}~\bibnamefont
  {M\"{u}hlbacher}}, \ and\ \bibinfo {author} {\bibfnamefont {C.}~\bibnamefont
  {Rentenberger}},\ }\href {\doibase
  https://doi.org/10.1016/1359-6454(95)00191-3} {\bibfield  {journal} {\bibinfo
   {journal} {Acta Materialia}\ }\textbf {\bibinfo {volume} {44}},\ \bibinfo
  {pages} {547} (\bibinfo {year} {1996})}\BibitemShut {NoStop}%
\bibitem [{\citenamefont {Umakoshi}\ \emph {et~al.}(1984)\citenamefont
  {Umakoshi}, \citenamefont {Pope},\ and\ \citenamefont
  {Vitek}}]{UMAKOSHI1984449}%
  \BibitemOpen
  \bibfield  {author} {\bibinfo {author} {\bibfnamefont {Y.}~\bibnamefont
  {Umakoshi}}, \bibinfo {author} {\bibfnamefont {D.}~\bibnamefont {Pope}}, \
  and\ \bibinfo {author} {\bibfnamefont {V.}~\bibnamefont {Vitek}},\ }\href
  {\doibase https://doi.org/10.1016/0001-6160(84)90118-4} {\bibfield  {journal}
  {\bibinfo  {journal} {Acta Metallurgica}\ }\textbf {\bibinfo {volume} {32}},\
  \bibinfo {pages} {449} (\bibinfo {year} {1984})}\BibitemShut {NoStop}%
\bibitem [{\citenamefont {Shang}\ \emph {et~al.}(2020)\citenamefont {Shang},
  \citenamefont {Shimanek}, \citenamefont {Qin}, \citenamefont {Wang},
  \citenamefont {Beese},\ and\ \citenamefont {Liu}}]{PhysRevB.101.024102}%
  \BibitemOpen
  \bibfield  {author} {\bibinfo {author} {\bibfnamefont {S.-L.}\ \bibnamefont
  {Shang}}, \bibinfo {author} {\bibfnamefont {J.}~\bibnamefont {Shimanek}},
  \bibinfo {author} {\bibfnamefont {S.}~\bibnamefont {Qin}}, \bibinfo {author}
  {\bibfnamefont {Y.}~\bibnamefont {Wang}}, \bibinfo {author} {\bibfnamefont
  {A.~M.}\ \bibnamefont {Beese}}, \ and\ \bibinfo {author} {\bibfnamefont
  {Z.-K.}\ \bibnamefont {Liu}},\ }\href {\doibase 10.1103/PhysRevB.101.024102}
  {\bibfield  {journal} {\bibinfo  {journal} {Phys. Rev. B}\ }\textbf {\bibinfo
  {volume} {101}},\ \bibinfo {pages} {024102} (\bibinfo {year}
  {2020})}\BibitemShut {NoStop}%
\bibitem [{\citenamefont {Schoeck}\ \emph {et~al.}(1999)\citenamefont
  {Schoeck}, \citenamefont {Kohlhammer},\ and\ \citenamefont
  {Fahnle}}]{doi10.1080/095008399176544}%
  \BibitemOpen
  \bibfield  {author} {\bibinfo {author} {\bibfnamefont {G.}~\bibnamefont
  {Schoeck}}, \bibinfo {author} {\bibfnamefont {S.}~\bibnamefont {Kohlhammer}},
  \ and\ \bibinfo {author} {\bibfnamefont {M.}~\bibnamefont {Fahnle}},\ }\href
  {\doibase 10.1080/095008399176544} {\bibfield  {journal} {\bibinfo  {journal}
  {Philosophical Magazine Letters}\ }\textbf {\bibinfo {volume} {79}},\
  \bibinfo {pages} {849} (\bibinfo {year} {1999})}\BibitemShut {NoStop}%
\bibitem [{\citenamefont {Wang-Koh}(2017)}]{doi10.1080/02670836.2016.1215961}%
  \BibitemOpen
  \bibfield  {author} {\bibinfo {author} {\bibfnamefont {Y.~M.}\ \bibnamefont
  {Wang-Koh}},\ }\href {\doibase 10.1080/02670836.2016.1215961} {\bibfield
  {journal} {\bibinfo  {journal} {Materials Science and Technology}\ }\textbf
  {\bibinfo {volume} {33}},\ \bibinfo {pages} {934} (\bibinfo {year}
  {2017})}\BibitemShut {NoStop}%
\bibitem [{\citenamefont {Abate}\ \emph {et~al.}(2015)\citenamefont {Abate},
  \citenamefont {Gamage}, \citenamefont {Zhen}, \citenamefont {Cronin},
  \citenamefont {Wang}, \citenamefont {Babicheva}, \citenamefont {H~Javani},\
  and\ \citenamefont {Stockman}}]{abate_nanoscopy_2016}%
  \BibitemOpen
  \bibfield  {author} {\bibinfo {author} {\bibfnamefont {Y.}~\bibnamefont
  {Abate}}, \bibinfo {author} {\bibfnamefont {S.}~\bibnamefont {Gamage}},
  \bibinfo {author} {\bibfnamefont {L.}~\bibnamefont {Zhen}}, \bibinfo {author}
  {\bibfnamefont {S.}~\bibnamefont {Cronin}}, \bibinfo {author} {\bibfnamefont
  {H.}~\bibnamefont {Wang}}, \bibinfo {author} {\bibfnamefont {V.}~\bibnamefont
  {Babicheva}}, \bibinfo {author} {\bibfnamefont {M.}~\bibnamefont {H~Javani}},
  \ and\ \bibinfo {author} {\bibfnamefont {M.}~\bibnamefont {Stockman}},\
  }\href {\doibase 10.1038/lsa.2016.162} {\bibfield  {journal} {\bibinfo
  {journal} {Light: Science \& Applications}\ }\textbf {\bibinfo {volume}
  {5}},\ \bibinfo {pages} {16162} (\bibinfo {year} {2015})}\BibitemShut
  {NoStop}%
\bibitem [{\citenamefont {Paidar}\ \emph {et~al.}(1984)\citenamefont {Paidar},
  \citenamefont {Pope},\ and\ \citenamefont {Vitek}}]{PAIDAR1984435}%
  \BibitemOpen
  \bibfield  {author} {\bibinfo {author} {\bibfnamefont {V.}~\bibnamefont
  {Paidar}}, \bibinfo {author} {\bibfnamefont {D.}~\bibnamefont {Pope}}, \ and\
  \bibinfo {author} {\bibfnamefont {V.}~\bibnamefont {Vitek}},\ }\href
  {\doibase https://doi.org/10.1016/0001-6160(84)90117-2} {\bibfield  {journal}
  {\bibinfo  {journal} {Acta Metallurgica}\ }\textbf {\bibinfo {volume} {32}},\
  \bibinfo {pages} {435} (\bibinfo {year} {1984})}\BibitemShut {NoStop}%
\bibitem [{\citenamefont {Chen}\ \emph {et~al.}(2022)\citenamefont {Chen},
  \citenamefont {Tamm}, \citenamefont {Wang}, \citenamefont {Epler},
  \citenamefont {Asta},\ and\ \citenamefont {Frolov}}]{chen_modeling_2022}%
  \BibitemOpen
  \bibfield  {author} {\bibinfo {author} {\bibfnamefont {E.}~\bibnamefont
  {Chen}}, \bibinfo {author} {\bibfnamefont {A.}~\bibnamefont {Tamm}}, \bibinfo
  {author} {\bibfnamefont {T.}~\bibnamefont {Wang}}, \bibinfo {author}
  {\bibfnamefont {M.~E.}\ \bibnamefont {Epler}}, \bibinfo {author}
  {\bibfnamefont {M.}~\bibnamefont {Asta}}, \ and\ \bibinfo {author}
  {\bibfnamefont {T.}~\bibnamefont {Frolov}},\ }\href {\doibase
  10.1038/s41524-022-00755-1} {\bibfield  {journal} {\bibinfo  {journal} {npj
  Computational Materials}\ }\textbf {\bibinfo {volume} {8}},\ \bibinfo {pages}
  {80} (\bibinfo {year} {2022})}\BibitemShut {NoStop}%
\bibitem [{\citenamefont {Tichy}\ \emph {et~al.}(1986)\citenamefont {Tichy},
  \citenamefont {Vitek},\ and\ \citenamefont
  {Pope}}]{doi10.1080/01418618608242846}%
  \BibitemOpen
  \bibfield  {author} {\bibinfo {author} {\bibfnamefont {G.}~\bibnamefont
  {Tichy}}, \bibinfo {author} {\bibfnamefont {V.}~\bibnamefont {Vitek}}, \ and\
  \bibinfo {author} {\bibfnamefont {D.~P.}\ \bibnamefont {Pope}},\ }\href
  {\doibase 10.1080/01418618608242846} {\bibfield  {journal} {\bibinfo
  {journal} {Philosophical Magazine A}\ }\textbf {\bibinfo {volume} {53}},\
  \bibinfo {pages} {467} (\bibinfo {year} {1986})}\BibitemShut {NoStop}%
\bibitem [{\citenamefont {Rae}\ and\ \citenamefont {Reed}(2007)}]{RAE20071067}%
  \BibitemOpen
  \bibfield  {author} {\bibinfo {author} {\bibfnamefont {C.}~\bibnamefont
  {Rae}}\ and\ \bibinfo {author} {\bibfnamefont {R.}~\bibnamefont {Reed}},\
  }\href {\doibase https://doi.org/10.1016/j.actamat.2006.09.026} {\bibfield
  {journal} {\bibinfo  {journal} {Acta Materialia}\ }\textbf {\bibinfo {volume}
  {55}},\ \bibinfo {pages} {1067} (\bibinfo {year} {2007})}\BibitemShut
  {NoStop}%
\bibitem [{\citenamefont {Matan}\ \emph {et~al.}(1999)\citenamefont {Matan},
  \citenamefont {Cox}, \citenamefont {Carter}, \citenamefont {Rist},
  \citenamefont {Rae},\ and\ \citenamefont {Reed}}]{MATAN19991549}%
  \BibitemOpen
  \bibfield  {author} {\bibinfo {author} {\bibfnamefont {N.}~\bibnamefont
  {Matan}}, \bibinfo {author} {\bibfnamefont {D.}~\bibnamefont {Cox}}, \bibinfo
  {author} {\bibfnamefont {P.}~\bibnamefont {Carter}}, \bibinfo {author}
  {\bibfnamefont {M.}~\bibnamefont {Rist}}, \bibinfo {author} {\bibfnamefont
  {C.}~\bibnamefont {Rae}}, \ and\ \bibinfo {author} {\bibfnamefont
  {R.}~\bibnamefont {Reed}},\ }\href {\doibase
  https://doi.org/10.1016/S1359-6454(99)00029-4} {\bibfield  {journal}
  {\bibinfo  {journal} {Acta Materialia}\ }\textbf {\bibinfo {volume} {47}},\
  \bibinfo {pages} {1549} (\bibinfo {year} {1999})}\BibitemShut {NoStop}%
\bibitem [{\citenamefont {Kear}\ \emph {et~al.}(1970)\citenamefont {Kear},
  \citenamefont {Oblak},\ and\ \citenamefont {Giamei}}]{kear_stacking_1970}%
  \BibitemOpen
  \bibfield  {author} {\bibinfo {author} {\bibfnamefont {B.~H.}\ \bibnamefont
  {Kear}}, \bibinfo {author} {\bibfnamefont {J.~M.}\ \bibnamefont {Oblak}}, \
  and\ \bibinfo {author} {\bibfnamefont {A.~F.}\ \bibnamefont {Giamei}},\
  }\href {\doibase 10.1007/BF03038373} {\bibfield  {journal} {\bibinfo
  {journal} {Metallurgical Transactions}\ }\textbf {\bibinfo {volume} {1}},\
  \bibinfo {pages} {2477} (\bibinfo {year} {1970})}\BibitemShut {NoStop}%
\bibitem [{\citenamefont {Leverant}\ \emph {et~al.}(1973)\citenamefont
  {Leverant}, \citenamefont {Kear},\ and\ \citenamefont
  {Oblak}}]{Leverant1973}%
  \BibitemOpen
  \bibfield  {author} {\bibinfo {author} {\bibfnamefont {G.~R.}\ \bibnamefont
  {Leverant}}, \bibinfo {author} {\bibfnamefont {B.~H.}\ \bibnamefont {Kear}},
  \ and\ \bibinfo {author} {\bibfnamefont {J.~M.}\ \bibnamefont {Oblak}},\
  }\href {\doibase 10.1007/BF02649637} {\bibfield  {journal} {\bibinfo
  {journal} {Metallurgical Transactions}\ }\textbf {\bibinfo {volume} {4}},\
  \bibinfo {pages} {355} (\bibinfo {year} {1973})}\BibitemShut {NoStop}%
\bibitem [{\citenamefont {Knowles}\ and\ \citenamefont
  {Chen}(2003)}]{KNOWLES200388}%
  \BibitemOpen
  \bibfield  {author} {\bibinfo {author} {\bibfnamefont {D.}~\bibnamefont
  {Knowles}}\ and\ \bibinfo {author} {\bibfnamefont {Q.}~\bibnamefont {Chen}},\
  }\href {\doibase https://doi.org/10.1016/S0921-5093(02)00172-7} {\bibfield
  {journal} {\bibinfo  {journal} {Materials Science and Engineering: A}\
  }\textbf {\bibinfo {volume} {340}},\ \bibinfo {pages} {88} (\bibinfo {year}
  {2003})}\BibitemShut {NoStop}%
\bibitem [{\citenamefont {Breidi}\ \emph {et~al.}(2018)\citenamefont {Breidi},
  \citenamefont {Allen},\ and\ \citenamefont {Mottura}}]{BREIDI201897}%
  \BibitemOpen
  \bibfield  {author} {\bibinfo {author} {\bibfnamefont {A.}~\bibnamefont
  {Breidi}}, \bibinfo {author} {\bibfnamefont {J.}~\bibnamefont {Allen}}, \
  and\ \bibinfo {author} {\bibfnamefont {A.}~\bibnamefont {Mottura}},\ }\href
  {\doibase https://doi.org/10.1016/j.actamat.2017.11.042} {\bibfield
  {journal} {\bibinfo  {journal} {Acta Materialia}\ }\textbf {\bibinfo {volume}
  {145}},\ \bibinfo {pages} {97} (\bibinfo {year} {2018})}\BibitemShut
  {NoStop}%
\bibitem [{\citenamefont {Rae}\ \emph {et~al.}(2001)\citenamefont {Rae},
  \citenamefont {Matan},\ and\ \citenamefont {Reed}}]{RAE2001125}%
  \BibitemOpen
  \bibfield  {author} {\bibinfo {author} {\bibfnamefont {C.}~\bibnamefont
  {Rae}}, \bibinfo {author} {\bibfnamefont {N.}~\bibnamefont {Matan}}, \ and\
  \bibinfo {author} {\bibfnamefont {R.}~\bibnamefont {Reed}},\ }\href {\doibase
  https://doi.org/10.1016/S0921-5093(00)01788-3} {\bibfield  {journal}
  {\bibinfo  {journal} {Materials Science and Engineering: A}\ }\textbf
  {\bibinfo {volume} {300}},\ \bibinfo {pages} {125} (\bibinfo {year}
  {2001})}\BibitemShut {NoStop}%
\bibitem [{\citenamefont {Kovarik}\ \emph {et~al.}(2009)\citenamefont
  {Kovarik}, \citenamefont {Unocic}, \citenamefont {Li}, \citenamefont
  {Sarosi}, \citenamefont {Shen}, \citenamefont {Wang},\ and\ \citenamefont
  {Mills}}]{KOVARIK2009839}%
  \BibitemOpen
  \bibfield  {author} {\bibinfo {author} {\bibfnamefont {L.}~\bibnamefont
  {Kovarik}}, \bibinfo {author} {\bibfnamefont {R.}~\bibnamefont {Unocic}},
  \bibinfo {author} {\bibfnamefont {J.}~\bibnamefont {Li}}, \bibinfo {author}
  {\bibfnamefont {P.}~\bibnamefont {Sarosi}}, \bibinfo {author} {\bibfnamefont
  {C.}~\bibnamefont {Shen}}, \bibinfo {author} {\bibfnamefont {Y.}~\bibnamefont
  {Wang}}, \ and\ \bibinfo {author} {\bibfnamefont {M.}~\bibnamefont {Mills}},\
  }\href {\doibase https://doi.org/10.1016/j.pmatsci.2009.03.010} {\bibfield
  {journal} {\bibinfo  {journal} {Progress in Materials Science}\ }\textbf
  {\bibinfo {volume} {54}},\ \bibinfo {pages} {839} (\bibinfo {year}
  {2009})}\BibitemShut {NoStop}%
\bibitem [{\citenamefont {Viswanathan}\ \emph {et~al.}(2005)\citenamefont
  {Viswanathan}, \citenamefont {Sarosi}, \citenamefont {Henry}, \citenamefont
  {Whitis}, \citenamefont {Milligan},\ and\ \citenamefont
  {Mills}}]{VISWANATHAN20053041}%
  \BibitemOpen
  \bibfield  {author} {\bibinfo {author} {\bibfnamefont {G.}~\bibnamefont
  {Viswanathan}}, \bibinfo {author} {\bibfnamefont {P.}~\bibnamefont {Sarosi}},
  \bibinfo {author} {\bibfnamefont {M.}~\bibnamefont {Henry}}, \bibinfo
  {author} {\bibfnamefont {D.}~\bibnamefont {Whitis}}, \bibinfo {author}
  {\bibfnamefont {W.}~\bibnamefont {Milligan}}, \ and\ \bibinfo {author}
  {\bibfnamefont {M.}~\bibnamefont {Mills}},\ }\href {\doibase
  https://doi.org/10.1016/j.actamat.2005.03.017} {\bibfield  {journal}
  {\bibinfo  {journal} {Acta Materialia}\ }\textbf {\bibinfo {volume} {53}},\
  \bibinfo {pages} {3041} (\bibinfo {year} {2005})}\BibitemShut {NoStop}%
\bibitem [{\citenamefont {Mryasov}\ \emph {et~al.}(2002)\citenamefont
  {Mryasov}, \citenamefont {Gornostyrev}, \citenamefont {van Schilfgaarde},\
  and\ \citenamefont {Freeman}}]{MRYASOV20024545}%
  \BibitemOpen
  \bibfield  {author} {\bibinfo {author} {\bibfnamefont {O.}~\bibnamefont
  {Mryasov}}, \bibinfo {author} {\bibfnamefont {Y.}~\bibnamefont
  {Gornostyrev}}, \bibinfo {author} {\bibfnamefont {M.}~\bibnamefont {van
  Schilfgaarde}}, \ and\ \bibinfo {author} {\bibfnamefont {A.}~\bibnamefont
  {Freeman}},\ }\href {\doibase https://doi.org/10.1016/S1359-6454(02)00282-3}
  {\bibfield  {journal} {\bibinfo  {journal} {Acta Materialia}\ }\textbf
  {\bibinfo {volume} {50}},\ \bibinfo {pages} {4545} (\bibinfo {year}
  {2002})}\BibitemShut {NoStop}%
\bibitem [{\citenamefont {Qi}\ and\ \citenamefont
  {Mishra}(2007)}]{PhysRevB.75.224105}%
  \BibitemOpen
  \bibfield  {author} {\bibinfo {author} {\bibfnamefont {Y.}~\bibnamefont
  {Qi}}\ and\ \bibinfo {author} {\bibfnamefont {R.~K.}\ \bibnamefont
  {Mishra}},\ }\href {\doibase 10.1103/PhysRevB.75.224105} {\bibfield
  {journal} {\bibinfo  {journal} {Phys. Rev. B}\ }\textbf {\bibinfo {volume}
  {75}},\ \bibinfo {pages} {224105} (\bibinfo {year} {2007})}\BibitemShut
  {NoStop}%
\bibitem [{\citenamefont {Vitos}\ \emph {et~al.}(2006)\citenamefont {Vitos},
  \citenamefont {Nilsson},\ and\ \citenamefont {Johansson}}]{VITOS20063821}%
  \BibitemOpen
  \bibfield  {author} {\bibinfo {author} {\bibfnamefont {L.}~\bibnamefont
  {Vitos}}, \bibinfo {author} {\bibfnamefont {J.-O.}\ \bibnamefont {Nilsson}},
  \ and\ \bibinfo {author} {\bibfnamefont {B.}~\bibnamefont {Johansson}},\
  }\href {\doibase https://doi.org/10.1016/j.actamat.2006.04.013} {\bibfield
  {journal} {\bibinfo  {journal} {Acta Materialia}\ }\textbf {\bibinfo {volume}
  {54}},\ \bibinfo {pages} {3821} (\bibinfo {year} {2006})}\BibitemShut
  {NoStop}%
\bibitem [{\citenamefont {Carter}\ and\ \citenamefont
  {Holmes}(1977)}]{doi10.1080/14786437708232942}%
  \BibitemOpen
  \bibfield  {author} {\bibinfo {author} {\bibfnamefont {C.~B.}\ \bibnamefont
  {Carter}}\ and\ \bibinfo {author} {\bibfnamefont {S.~M.}\ \bibnamefont
  {Holmes}},\ }\href {\doibase 10.1080/14786437708232942} {\bibfield  {journal}
  {\bibinfo  {journal} {The Philosophical Magazine: A Journal of Theoretical
  Experimental and Applied Physics}\ }\textbf {\bibinfo {volume} {35}},\
  \bibinfo {pages} {1161} (\bibinfo {year} {1977})}\BibitemShut {NoStop}%
\bibitem [{\citenamefont {de~Campos}(2008)}]{decampos2008}%
  \BibitemOpen
  \bibfield  {author} {\bibinfo {author} {\bibfnamefont {M.~F.}\ \bibnamefont
  {de~Campos}},\ }in\ \href {\doibase
  10.4028/www.scientific.net/MSF.591-593.708} {\emph {\bibinfo {booktitle}
  {Advanced Powder Technology VI}}},\ \bibinfo {series} {Materials Science
  Forum}, Vol.\ \bibinfo {volume} {591}\ (\bibinfo  {publisher} {Trans Tech
  Publications Ltd},\ \bibinfo {year} {2008})\ pp.\ \bibinfo {pages}
  {708--711}\BibitemShut {NoStop}%
\bibitem [{\citenamefont {Ma}\ \emph {et~al.}(2007)\citenamefont {Ma},
  \citenamefont {Carroll},\ and\ \citenamefont {Pollock}}]{MA20075802}%
  \BibitemOpen
  \bibfield  {author} {\bibinfo {author} {\bibfnamefont {S.}~\bibnamefont
  {Ma}}, \bibinfo {author} {\bibfnamefont {L.}~\bibnamefont {Carroll}}, \ and\
  \bibinfo {author} {\bibfnamefont {T.}~\bibnamefont {Pollock}},\ }\href
  {\doibase https://doi.org/10.1016/j.actamat.2007.06.042} {\bibfield
  {journal} {\bibinfo  {journal} {Acta Materialia}\ }\textbf {\bibinfo {volume}
  {55}},\ \bibinfo {pages} {5802} (\bibinfo {year} {2007})}\BibitemShut
  {NoStop}%
\bibitem [{\citenamefont {Gallagher}(1970)}]{gallagher_influence_1970}%
  \BibitemOpen
  \bibfield  {author} {\bibinfo {author} {\bibfnamefont {P.~C.~J.}\
  \bibnamefont {Gallagher}},\ }\href {\doibase 10.1007/BF03038370} {\bibfield
  {journal} {\bibinfo  {journal} {Metallurgical Transactions}\ }\textbf
  {\bibinfo {volume} {1}},\ \bibinfo {pages} {2429} (\bibinfo {year}
  {1970})}\BibitemShut {NoStop}%
\bibitem [{\citenamefont {Xie}\ \emph {et~al.}(1982)\citenamefont {Xie},
  \citenamefont {Chen}, \citenamefont {McHugh},\ and\ \citenamefont
  {Tien}}]{XIE1982483}%
  \BibitemOpen
  \bibfield  {author} {\bibinfo {author} {\bibfnamefont {X.}~\bibnamefont
  {Xie}}, \bibinfo {author} {\bibfnamefont {G.}~\bibnamefont {Chen}}, \bibinfo
  {author} {\bibfnamefont {P.}~\bibnamefont {McHugh}}, \ and\ \bibinfo {author}
  {\bibfnamefont {J.}~\bibnamefont {Tien}},\ }\href {\doibase
  https://doi.org/10.1016/0036-9748(82)90254-X} {\bibfield  {journal} {\bibinfo
   {journal} {Scripta Metallurgica}\ }\textbf {\bibinfo {volume} {16}},\
  \bibinfo {pages} {483} (\bibinfo {year} {1982})}\BibitemShut {NoStop}%
\bibitem [{\citenamefont {Li}\ \emph {et~al.}(2020)\citenamefont {Li},
  \citenamefont {Xie}, \citenamefont {Wang}, \citenamefont {Liaw},\ and\
  \citenamefont {Zhang}}]{10.3389/fmats.2020.00290}%
  \BibitemOpen
  \bibfield  {author} {\bibinfo {author} {\bibfnamefont {R.}~\bibnamefont
  {Li}}, \bibinfo {author} {\bibfnamefont {L.}~\bibnamefont {Xie}}, \bibinfo
  {author} {\bibfnamefont {W.~Y.}\ \bibnamefont {Wang}}, \bibinfo {author}
  {\bibfnamefont {P.~K.}\ \bibnamefont {Liaw}}, \ and\ \bibinfo {author}
  {\bibfnamefont {Y.}~\bibnamefont {Zhang}},\ }\href {\doibase
  10.3389/fmats.2020.00290} {\bibfield  {journal} {\bibinfo  {journal}
  {Frontiers in Materials}\ }\textbf {\bibinfo {volume} {7}} (\bibinfo {year}
  {2020}),\ 10.3389/fmats.2020.00290}\BibitemShut {NoStop}%
\bibitem [{\citenamefont {Ikeda}\ \emph {et~al.}(2019)\citenamefont {Ikeda},
  \citenamefont {Grabowski},\ and\ \citenamefont {K\"{o}rmann}}]{IKEDA2019464}%
  \BibitemOpen
  \bibfield  {author} {\bibinfo {author} {\bibfnamefont {Y.}~\bibnamefont
  {Ikeda}}, \bibinfo {author} {\bibfnamefont {B.}~\bibnamefont {Grabowski}}, \
  and\ \bibinfo {author} {\bibfnamefont {F.}~\bibnamefont {K\"{o}rmann}},\
  }\href {\doibase https://doi.org/10.1016/j.matchar.2018.06.019} {\bibfield
  {journal} {\bibinfo  {journal} {Materials Characterization}\ }\textbf
  {\bibinfo {volume} {147}},\ \bibinfo {pages} {464} (\bibinfo {year}
  {2019})}\BibitemShut {NoStop}%
\bibitem [{\citenamefont {Curtarolo}\ \emph {et~al.}(2013)\citenamefont
  {Curtarolo}, \citenamefont {Hart}, \citenamefont {Nardelli}, \citenamefont
  {Mingo}, \citenamefont {Sanvito},\ and\ \citenamefont
  {Levy}}]{Curtarolo2013}%
  \BibitemOpen
  \bibfield  {author} {\bibinfo {author} {\bibfnamefont {S.}~\bibnamefont
  {Curtarolo}}, \bibinfo {author} {\bibfnamefont {G.~L.~W.}\ \bibnamefont
  {Hart}}, \bibinfo {author} {\bibfnamefont {M.~B.}\ \bibnamefont {Nardelli}},
  \bibinfo {author} {\bibfnamefont {N.}~\bibnamefont {Mingo}}, \bibinfo
  {author} {\bibfnamefont {S.}~\bibnamefont {Sanvito}}, \ and\ \bibinfo
  {author} {\bibfnamefont {O.}~\bibnamefont {Levy}},\ }\href {\doibase
  10.1038/nmat3568} {\bibfield  {journal} {\bibinfo  {journal} {Nature
  Materials}\ }\textbf {\bibinfo {volume} {12}},\ \bibinfo {pages} {191}
  (\bibinfo {year} {2013})}\BibitemShut {NoStop}%
\bibitem [{\citenamefont {Hart}\ \emph {et~al.}(2021)\citenamefont {Hart},
  \citenamefont {Mueller}, \citenamefont {Toher},\ and\ \citenamefont
  {Curtarolo}}]{Hart2021}%
  \BibitemOpen
  \bibfield  {author} {\bibinfo {author} {\bibfnamefont {G.~L.~W.}\
  \bibnamefont {Hart}}, \bibinfo {author} {\bibfnamefont {T.}~\bibnamefont
  {Mueller}}, \bibinfo {author} {\bibfnamefont {C.}~\bibnamefont {Toher}}, \
  and\ \bibinfo {author} {\bibfnamefont {S.}~\bibnamefont {Curtarolo}},\ }\href
  {\doibase 10.1038/s41578-021-00340-w} {\bibfield  {journal} {\bibinfo
  {journal} {Nature Reviews Materials}\ }\textbf {\bibinfo {volume} {6}},\
  \bibinfo {pages} {730} (\bibinfo {year} {2021})}\BibitemShut {NoStop}%
\bibitem [{\citenamefont {V\'{i}tek}(1968)}]{doi10.1080/14786436808227500}%
  \BibitemOpen
  \bibfield  {author} {\bibinfo {author} {\bibfnamefont {V.}~\bibnamefont
  {V\'{i}tek}},\ }\href {\doibase 10.1080/14786436808227500} {\bibfield
  {journal} {\bibinfo  {journal} {The Philosophical Magazine: A Journal of
  Theoretical Experimental and Applied Physics}\ }\textbf {\bibinfo {volume}
  {18}},\ \bibinfo {pages} {773} (\bibinfo {year} {1968})}\BibitemShut
  {NoStop}%
\bibitem [{\citenamefont {Li}\ \emph {et~al.}(2009)\citenamefont {Li},
  \citenamefont {Ung\'{a}r}, \citenamefont {Wang}, \citenamefont {Morris},
  \citenamefont {Tichy}, \citenamefont {Lendvai}, \citenamefont {Yang},
  \citenamefont {Ren}, \citenamefont {Choo},\ and\ \citenamefont
  {Liaw}}]{LI20094988}%
  \BibitemOpen
  \bibfield  {author} {\bibinfo {author} {\bibfnamefont {L.}~\bibnamefont
  {Li}}, \bibinfo {author} {\bibfnamefont {T.}~\bibnamefont {Ung\'{a}r}},
  \bibinfo {author} {\bibfnamefont {Y.}~\bibnamefont {Wang}}, \bibinfo {author}
  {\bibfnamefont {J.}~\bibnamefont {Morris}}, \bibinfo {author} {\bibfnamefont
  {G.}~\bibnamefont {Tichy}}, \bibinfo {author} {\bibfnamefont
  {J.}~\bibnamefont {Lendvai}}, \bibinfo {author} {\bibfnamefont
  {Y.}~\bibnamefont {Yang}}, \bibinfo {author} {\bibfnamefont {Y.}~\bibnamefont
  {Ren}}, \bibinfo {author} {\bibfnamefont {H.}~\bibnamefont {Choo}}, \ and\
  \bibinfo {author} {\bibfnamefont {P.}~\bibnamefont {Liaw}},\ }\href {\doibase
  https://doi.org/10.1016/j.actamat.2009.07.002} {\bibfield  {journal}
  {\bibinfo  {journal} {Acta Materialia}\ }\textbf {\bibinfo {volume} {57}},\
  \bibinfo {pages} {4988} (\bibinfo {year} {2009})}\BibitemShut {NoStop}%
\bibitem [{\citenamefont {Pierce}\ \emph {et~al.}(2015)\citenamefont {Pierce},
  \citenamefont {Jim\'{e}nez}, \citenamefont {Bentley}, \citenamefont {Raabe},\
  and\ \citenamefont {Wittig}}]{PIERCE2015178}%
  \BibitemOpen
  \bibfield  {author} {\bibinfo {author} {\bibfnamefont {D.}~\bibnamefont
  {Pierce}}, \bibinfo {author} {\bibfnamefont {J.}~\bibnamefont {Jim\'{e}nez}},
  \bibinfo {author} {\bibfnamefont {J.}~\bibnamefont {Bentley}}, \bibinfo
  {author} {\bibfnamefont {D.}~\bibnamefont {Raabe}}, \ and\ \bibinfo {author}
  {\bibfnamefont {J.}~\bibnamefont {Wittig}},\ }\href {\doibase
  https://doi.org/10.1016/j.actamat.2015.08.030} {\bibfield  {journal}
  {\bibinfo  {journal} {Acta Materialia}\ }\textbf {\bibinfo {volume} {100}},\
  \bibinfo {pages} {178} (\bibinfo {year} {2015})}\BibitemShut {NoStop}%
\bibitem [{\citenamefont {Tadmor}\ and\ \citenamefont
  {Bernstein}(2004)}]{TADMOR20042507}%
  \BibitemOpen
  \bibfield  {author} {\bibinfo {author} {\bibfnamefont {E.}~\bibnamefont
  {Tadmor}}\ and\ \bibinfo {author} {\bibfnamefont {N.}~\bibnamefont
  {Bernstein}},\ }\href {\doibase https://doi.org/10.1016/j.jmps.2004.05.002}
  {\bibfield  {journal} {\bibinfo  {journal} {Journal of the Mechanics and
  Physics of Solids}\ }\textbf {\bibinfo {volume} {52}},\ \bibinfo {pages}
  {2507} (\bibinfo {year} {2004})}\BibitemShut {NoStop}%
\bibitem [{\citenamefont {Cahoon}\ \emph {et~al.}(2009)\citenamefont {Cahoon},
  \citenamefont {Li},\ and\ \citenamefont {Richards}}]{CAHOON200956}%
  \BibitemOpen
  \bibfield  {author} {\bibinfo {author} {\bibfnamefont {J.}~\bibnamefont
  {Cahoon}}, \bibinfo {author} {\bibfnamefont {Q.}~\bibnamefont {Li}}, \ and\
  \bibinfo {author} {\bibfnamefont {N.}~\bibnamefont {Richards}},\ }\href
  {\doibase https://doi.org/10.1016/j.msea.2009.07.021} {\bibfield  {journal}
  {\bibinfo  {journal} {Materials Science and Engineering: A}\ }\textbf
  {\bibinfo {volume} {526}},\ \bibinfo {pages} {56} (\bibinfo {year}
  {2009})}\BibitemShut {NoStop}%
\bibitem [{\citenamefont {Sarma}\ \emph {et~al.}(2010)\citenamefont {Sarma},
  \citenamefont {Wang}, \citenamefont {Jian}, \citenamefont {Kauffmann},
  \citenamefont {Conrad}, \citenamefont {Freudenberger},\ and\ \citenamefont
  {Zhu}}]{SARMA20107624}%
  \BibitemOpen
  \bibfield  {author} {\bibinfo {author} {\bibfnamefont {V.~S.}\ \bibnamefont
  {Sarma}}, \bibinfo {author} {\bibfnamefont {J.}~\bibnamefont {Wang}},
  \bibinfo {author} {\bibfnamefont {W.}~\bibnamefont {Jian}}, \bibinfo {author}
  {\bibfnamefont {A.}~\bibnamefont {Kauffmann}}, \bibinfo {author}
  {\bibfnamefont {H.}~\bibnamefont {Conrad}}, \bibinfo {author} {\bibfnamefont
  {J.}~\bibnamefont {Freudenberger}}, \ and\ \bibinfo {author} {\bibfnamefont
  {Y.}~\bibnamefont {Zhu}},\ }\href {\doibase
  https://doi.org/10.1016/j.msea.2010.08.015} {\bibfield  {journal} {\bibinfo
  {journal} {Materials Science and Engineering: A}\ }\textbf {\bibinfo {volume}
  {527}},\ \bibinfo {pages} {7624} (\bibinfo {year} {2010})}\BibitemShut
  {NoStop}%
\bibitem [{\citenamefont {Chandran}\ and\ \citenamefont
  {Sondhi}(2011)}]{doi10.1063/1.3585786}%
  \BibitemOpen
  \bibfield  {author} {\bibinfo {author} {\bibfnamefont {M.}~\bibnamefont
  {Chandran}}\ and\ \bibinfo {author} {\bibfnamefont {S.~K.}\ \bibnamefont
  {Sondhi}},\ }\href {\doibase 10.1063/1.3585786} {\bibfield  {journal}
  {\bibinfo  {journal} {Journal of Applied Physics}\ }\textbf {\bibinfo
  {volume} {109}},\ \bibinfo {pages} {103525} (\bibinfo {year}
  {2011})}\BibitemShut {NoStop}%
\bibitem [{\citenamefont {Mohamed}\ and\ \citenamefont
  {Langdon}(1974)}]{MOHAMED1974779}%
  \BibitemOpen
  \bibfield  {author} {\bibinfo {author} {\bibfnamefont {F.~A.}\ \bibnamefont
  {Mohamed}}\ and\ \bibinfo {author} {\bibfnamefont {T.~G.}\ \bibnamefont
  {Langdon}},\ }\href {\doibase https://doi.org/10.1016/0001-6160(74)90088-1}
  {\bibfield  {journal} {\bibinfo  {journal} {Acta Metallurgica}\ }\textbf
  {\bibinfo {volume} {22}},\ \bibinfo {pages} {779} (\bibinfo {year}
  {1974})}\BibitemShut {NoStop}%
\bibitem [{\citenamefont {Argon}\ and\ \citenamefont
  {Moffatt}(1981)}]{ARGON1981293}%
  \BibitemOpen
  \bibfield  {author} {\bibinfo {author} {\bibfnamefont {A.}~\bibnamefont
  {Argon}}\ and\ \bibinfo {author} {\bibfnamefont {W.}~\bibnamefont
  {Moffatt}},\ }\href {\doibase https://doi.org/10.1016/0001-6160(81)90156-5}
  {\bibfield  {journal} {\bibinfo  {journal} {Acta Metallurgica}\ }\textbf
  {\bibinfo {volume} {29}},\ \bibinfo {pages} {293} (\bibinfo {year}
  {1981})}\BibitemShut {NoStop}%
\bibitem [{\citenamefont {Yu}\ and\ \citenamefont {Wang}(2009)}]{YU20095914}%
  \BibitemOpen
  \bibfield  {author} {\bibinfo {author} {\bibfnamefont {X.-X.}\ \bibnamefont
  {Yu}}\ and\ \bibinfo {author} {\bibfnamefont {C.-Y.}\ \bibnamefont {Wang}},\
  }\href {\doibase https://doi.org/10.1016/j.actamat.2009.08.019} {\bibfield
  {journal} {\bibinfo  {journal} {Acta Materialia}\ }\textbf {\bibinfo {volume}
  {57}},\ \bibinfo {pages} {5914} (\bibinfo {year} {2009})}\BibitemShut
  {NoStop}%
\bibitem [{\citenamefont {Shang}\ \emph
  {et~al.}(2012{\natexlab{a}})\citenamefont {Shang}, \citenamefont {Zacherl},
  \citenamefont {Fang}, \citenamefont {Wang}, \citenamefont {Y.Du},\ and\
  \citenamefont {Liu}}]{Shang_2012}%
  \BibitemOpen
  \bibfield  {author} {\bibinfo {author} {\bibfnamefont {S.}~\bibnamefont
  {Shang}}, \bibinfo {author} {\bibfnamefont {C.}~\bibnamefont {Zacherl}},
  \bibinfo {author} {\bibfnamefont {H.}~\bibnamefont {Fang}}, \bibinfo {author}
  {\bibfnamefont {Y.}~\bibnamefont {Wang}}, \bibinfo {author} {\bibnamefont
  {Y.Du}}, \ and\ \bibinfo {author} {\bibfnamefont {Z.}~\bibnamefont {Liu}},\
  }\href {\doibase 10.1088/0953-8984/24/50/505403} {\bibfield  {journal}
  {\bibinfo  {journal} {Journal of Physics: Condensed Matter}\ }\textbf
  {\bibinfo {volume} {24}},\ \bibinfo {pages} {505403} (\bibinfo {year}
  {2012}{\natexlab{a}})}\BibitemShut {NoStop}%
\bibitem [{\citenamefont {Shang}\ \emph
  {et~al.}(2012{\natexlab{b}})\citenamefont {Shang}, \citenamefont {Wang},
  \citenamefont {Wang}, \citenamefont {Du}, \citenamefont {Zhang},
  \citenamefont {Patel},\ and\ \citenamefont {Liu}}]{Shang_2012_2}%
  \BibitemOpen
  \bibfield  {author} {\bibinfo {author} {\bibfnamefont {S.}~\bibnamefont
  {Shang}}, \bibinfo {author} {\bibfnamefont {W.}~\bibnamefont {Wang}},
  \bibinfo {author} {\bibfnamefont {Y.}~\bibnamefont {Wang}}, \bibinfo {author}
  {\bibfnamefont {Y.}~\bibnamefont {Du}}, \bibinfo {author} {\bibfnamefont
  {J.}~\bibnamefont {Zhang}}, \bibinfo {author} {\bibfnamefont
  {A.}~\bibnamefont {Patel}}, \ and\ \bibinfo {author} {\bibfnamefont
  {Z.}~\bibnamefont {Liu}},\ }\href {\doibase 10.1088/0953-8984/24/15/155402}
  {\bibfield  {journal} {\bibinfo  {journal} {Journal of Physics: Condensed
  Matter}\ }\textbf {\bibinfo {volume} {24}},\ \bibinfo {pages} {155402}
  (\bibinfo {year} {2012}{\natexlab{b}})}\BibitemShut {NoStop}%
\bibitem [{\citenamefont {Siegel}(2005)}]{doi10.1063/1.2051793}%
  \BibitemOpen
  \bibfield  {author} {\bibinfo {author} {\bibfnamefont {D.~J.}\ \bibnamefont
  {Siegel}},\ }\href {\doibase 10.1063/1.2051793} {\bibfield  {journal}
  {\bibinfo  {journal} {Applied Physics Letters}\ }\textbf {\bibinfo {volume}
  {87}},\ \bibinfo {pages} {121901} (\bibinfo {year} {2005})}\BibitemShut
  {NoStop}%
\bibitem [{\citenamefont {Yu}\ and\ \citenamefont {Wang}(2012)}]{YU201238}%
  \BibitemOpen
  \bibfield  {author} {\bibinfo {author} {\bibfnamefont {X.-X.}\ \bibnamefont
  {Yu}}\ and\ \bibinfo {author} {\bibfnamefont {C.-Y.}\ \bibnamefont {Wang}},\
  }\href {\doibase https://doi.org/10.1016/j.msea.2011.12.112} {\bibfield
  {journal} {\bibinfo  {journal} {Materials Science and Engineering: A}\
  }\textbf {\bibinfo {volume} {539}},\ \bibinfo {pages} {38} (\bibinfo {year}
  {2012})}\BibitemShut {NoStop}%
\bibitem [{\citenamefont {Eurich}\ and\ \citenamefont
  {Bristowe}(2015)}]{EURICH201587}%
  \BibitemOpen
  \bibfield  {author} {\bibinfo {author} {\bibfnamefont {N.}~\bibnamefont
  {Eurich}}\ and\ \bibinfo {author} {\bibfnamefont {P.}~\bibnamefont
  {Bristowe}},\ }\href {\doibase
  https://doi.org/10.1016/j.scriptamat.2015.02.020} {\bibfield  {journal}
  {\bibinfo  {journal} {Scripta Materialia}\ }\textbf {\bibinfo {volume}
  {102}},\ \bibinfo {pages} {87} (\bibinfo {year} {2015})}\BibitemShut
  {NoStop}%
\bibitem [{\citenamefont {Yang}\ \emph {et~al.}(2020)\citenamefont {Yang},
  \citenamefont {Qu}, \citenamefont {Sun}, \citenamefont {Yue}, \citenamefont
  {Su}, \citenamefont {Zhang},\ and\ \citenamefont {Liu}}]{YANG2020109682}%
  \BibitemOpen
  \bibfield  {author} {\bibinfo {author} {\bibfnamefont {W.}~\bibnamefont
  {Yang}}, \bibinfo {author} {\bibfnamefont {P.}~\bibnamefont {Qu}}, \bibinfo
  {author} {\bibfnamefont {J.}~\bibnamefont {Sun}}, \bibinfo {author}
  {\bibfnamefont {Q.}~\bibnamefont {Yue}}, \bibinfo {author} {\bibfnamefont
  {H.}~\bibnamefont {Su}}, \bibinfo {author} {\bibfnamefont {J.}~\bibnamefont
  {Zhang}}, \ and\ \bibinfo {author} {\bibfnamefont {L.}~\bibnamefont {Liu}},\
  }\href {\doibase https://doi.org/10.1016/j.vacuum.2020.109682} {\bibfield
  {journal} {\bibinfo  {journal} {Vacuum}\ }\textbf {\bibinfo {volume} {181}},\
  \bibinfo {pages} {109682} (\bibinfo {year} {2020})}\BibitemShut {NoStop}%
\bibitem [{\citenamefont {Hu}\ \emph {et~al.}(2020)\citenamefont {Hu},
  \citenamefont {Zhang}, \citenamefont {Chen}, \citenamefont {He},\ and\
  \citenamefont {Guo}}]{HU2020155799}%
  \BibitemOpen
  \bibfield  {author} {\bibinfo {author} {\bibfnamefont {C.}~\bibnamefont
  {Hu}}, \bibinfo {author} {\bibfnamefont {Z.}~\bibnamefont {Zhang}}, \bibinfo
  {author} {\bibfnamefont {H.}~\bibnamefont {Chen}}, \bibinfo {author}
  {\bibfnamefont {J.}~\bibnamefont {He}}, \ and\ \bibinfo {author}
  {\bibfnamefont {H.}~\bibnamefont {Guo}},\ }\href {\doibase
  https://doi.org/10.1016/j.jallcom.2020.155799} {\bibfield  {journal}
  {\bibinfo  {journal} {Journal of Alloys and Compounds}\ }\textbf {\bibinfo
  {volume} {843}},\ \bibinfo {pages} {155799} (\bibinfo {year}
  {2020})}\BibitemShut {NoStop}%
\bibitem [{\citenamefont {Xia}\ \emph {et~al.}(2022)\citenamefont {Xia},
  \citenamefont {Xu}, \citenamefont {Shi}, \citenamefont {Xie},\ and\
  \citenamefont {Chen}}]{XIA2022104183}%
  \BibitemOpen
  \bibfield  {author} {\bibinfo {author} {\bibfnamefont {F.}~\bibnamefont
  {Xia}}, \bibinfo {author} {\bibfnamefont {W.}~\bibnamefont {Xu}}, \bibinfo
  {author} {\bibfnamefont {Z.}~\bibnamefont {Shi}}, \bibinfo {author}
  {\bibfnamefont {W.}~\bibnamefont {Xie}}, \ and\ \bibinfo {author}
  {\bibfnamefont {L.}~\bibnamefont {Chen}},\ }\href {\doibase
  https://doi.org/10.1016/j.mechmat.2021.104183} {\bibfield  {journal}
  {\bibinfo  {journal} {Mechanics of Materials}\ }\textbf {\bibinfo {volume}
  {165}},\ \bibinfo {pages} {104183} (\bibinfo {year} {2022})}\BibitemShut
  {NoStop}%
\bibitem [{\citenamefont {Zhao}\ \emph {et~al.}(2022)\citenamefont {Zhao},
  \citenamefont {Wang}, \citenamefont {Song}, \citenamefont {Wang},\ and\
  \citenamefont {Chen}}]{ZHAO2022110990}%
  \BibitemOpen
  \bibfield  {author} {\bibinfo {author} {\bibfnamefont {X.}~\bibnamefont
  {Zhao}}, \bibinfo {author} {\bibfnamefont {Y.}~\bibnamefont {Wang}}, \bibinfo
  {author} {\bibfnamefont {X.}~\bibnamefont {Song}}, \bibinfo {author}
  {\bibfnamefont {Y.}~\bibnamefont {Wang}}, \ and\ \bibinfo {author}
  {\bibfnamefont {Z.}~\bibnamefont {Chen}},\ }\href {\doibase
  https://doi.org/10.1016/j.commatsci.2021.110990} {\bibfield  {journal}
  {\bibinfo  {journal} {Computational Materials Science}\ }\textbf {\bibinfo
  {volume} {202}},\ \bibinfo {pages} {110990} (\bibinfo {year}
  {2022})}\BibitemShut {NoStop}%
\bibitem [{\citenamefont {Kresse}\ and\ \citenamefont
  {Furthm\"{u}ller}(1996{\natexlab{a}})}]{Kresse1996Efficiency}%
  \BibitemOpen
  \bibfield  {author} {\bibinfo {author} {\bibfnamefont {G.}~\bibnamefont
  {Kresse}}\ and\ \bibinfo {author} {\bibfnamefont {J.}~\bibnamefont
  {Furthm\"{u}ller}},\ }\href {\doibase
  https://doi.org/10.1016/0927-0256(96)00008-0} {\bibfield  {journal} {\bibinfo
   {journal} {Computational Materials Science}\ }\textbf {\bibinfo {volume}
  {6}},\ \bibinfo {pages} {15} (\bibinfo {year}
  {1996}{\natexlab{a}})}\BibitemShut {NoStop}%
\bibitem [{\citenamefont {Kresse}\ and\ \citenamefont
  {Furthm\"{u}ller}(1996{\natexlab{b}})}]{PhysRevB.54.11169}%
  \BibitemOpen
  \bibfield  {author} {\bibinfo {author} {\bibfnamefont {G.}~\bibnamefont
  {Kresse}}\ and\ \bibinfo {author} {\bibfnamefont {J.}~\bibnamefont
  {Furthm\"{u}ller}},\ }\href {\doibase 10.1103/PhysRevB.54.11169} {\bibfield
  {journal} {\bibinfo  {journal} {Phys. Rev. B}\ }\textbf {\bibinfo {volume}
  {54}},\ \bibinfo {pages} {11169} (\bibinfo {year}
  {1996}{\natexlab{b}})}\BibitemShut {NoStop}%
\bibitem [{\citenamefont {Perdew}\ \emph {et~al.}(1996)\citenamefont {Perdew},
  \citenamefont {Burke},\ and\ \citenamefont
  {Ernzerhof}}]{PhysRevLett.77.3865}%
  \BibitemOpen
  \bibfield  {author} {\bibinfo {author} {\bibfnamefont {J.~P.}\ \bibnamefont
  {Perdew}}, \bibinfo {author} {\bibfnamefont {K.}~\bibnamefont {Burke}}, \
  and\ \bibinfo {author} {\bibfnamefont {M.}~\bibnamefont {Ernzerhof}},\ }\href
  {\doibase 10.1103/PhysRevLett.77.3865} {\bibfield  {journal} {\bibinfo
  {journal} {Phys. Rev. Lett.}\ }\textbf {\bibinfo {volume} {77}},\ \bibinfo
  {pages} {3865} (\bibinfo {year} {1996})}\BibitemShut {NoStop}%
\bibitem [{\citenamefont {Methfessel}\ and\ \citenamefont
  {Paxton}(1989)}]{PhysRevB.40.3616}%
  \BibitemOpen
  \bibfield  {author} {\bibinfo {author} {\bibfnamefont {M.}~\bibnamefont
  {Methfessel}}\ and\ \bibinfo {author} {\bibfnamefont {A.}~\bibnamefont
  {Paxton}},\ }\href {\doibase 10.1103/PhysRevB.40.3616} {\bibfield  {journal}
  {\bibinfo  {journal} {Phys. Rev. B}\ }\textbf {\bibinfo {volume} {40}},\
  \bibinfo {pages} {3616} (\bibinfo {year} {1989})}\BibitemShut {NoStop}%
\bibitem [{\citenamefont {Bl\"{o}chl}\ \emph {et~al.}(1994)\citenamefont
  {Bl\"{o}chl}, \citenamefont {Jepsen},\ and\ \citenamefont
  {Andersen}}]{PhysRevB.49.16223}%
  \BibitemOpen
  \bibfield  {author} {\bibinfo {author} {\bibfnamefont {P.~E.}\ \bibnamefont
  {Bl\"{o}chl}}, \bibinfo {author} {\bibfnamefont {O.}~\bibnamefont {Jepsen}},
  \ and\ \bibinfo {author} {\bibfnamefont {O.~K.}\ \bibnamefont {Andersen}},\
  }\href {\doibase 10.1103/PhysRevB.49.16223} {\bibfield  {journal} {\bibinfo
  {journal} {Phys. Rev. B}\ }\textbf {\bibinfo {volume} {49}},\ \bibinfo
  {pages} {16223} (\bibinfo {year} {1994})}\BibitemShut {NoStop}%
\bibitem [{\citenamefont {Datta}\ \emph {et~al.}(2009)\citenamefont {Datta},
  \citenamefont {Waghmare},\ and\ \citenamefont {Ramamurty}}]{DATTA2009124}%
  \BibitemOpen
  \bibfield  {author} {\bibinfo {author} {\bibfnamefont {A.}~\bibnamefont
  {Datta}}, \bibinfo {author} {\bibfnamefont {U.}~\bibnamefont {Waghmare}}, \
  and\ \bibinfo {author} {\bibfnamefont {U.}~\bibnamefont {Ramamurty}},\ }\href
  {\doibase https://doi.org/10.1016/j.scriptamat.2008.09.018} {\bibfield
  {journal} {\bibinfo  {journal} {Scripta Materialia}\ }\textbf {\bibinfo
  {volume} {60}},\ \bibinfo {pages} {124} (\bibinfo {year} {2009})}\BibitemShut
  {NoStop}%
\bibitem [{\citenamefont {Han}\ \emph {et~al.}(2011)\citenamefont {Han},
  \citenamefont {Su}, \citenamefont {Jin},\ and\ \citenamefont
  {Zhu}}]{HAN2011693}%
  \BibitemOpen
  \bibfield  {author} {\bibinfo {author} {\bibfnamefont {J.}~\bibnamefont
  {Han}}, \bibinfo {author} {\bibfnamefont {X.}~\bibnamefont {Su}}, \bibinfo
  {author} {\bibfnamefont {Z.-H.}\ \bibnamefont {Jin}}, \ and\ \bibinfo
  {author} {\bibfnamefont {Y.}~\bibnamefont {Zhu}},\ }\href {\doibase
  https://doi.org/10.1016/j.scriptamat.2010.11.034} {\bibfield  {journal}
  {\bibinfo  {journal} {Scripta Materialia}\ }\textbf {\bibinfo {volume}
  {64}},\ \bibinfo {pages} {693} (\bibinfo {year} {2011})}\BibitemShut
  {NoStop}%
\bibitem [{\citenamefont {Ogata}\ \emph {et~al.}(2002)\citenamefont {Ogata},
  \citenamefont {Li},\ and\ \citenamefont {Yip}}]{doi10.1126/science.1076652}%
  \BibitemOpen
  \bibfield  {author} {\bibinfo {author} {\bibfnamefont {S.}~\bibnamefont
  {Ogata}}, \bibinfo {author} {\bibfnamefont {J.}~\bibnamefont {Li}}, \ and\
  \bibinfo {author} {\bibfnamefont {S.}~\bibnamefont {Yip}},\ }\href {\doibase
  10.1126/science.1076652} {\bibfield  {journal} {\bibinfo  {journal}
  {Science}\ }\textbf {\bibinfo {volume} {298}},\ \bibinfo {pages} {807}
  (\bibinfo {year} {2002})}\BibitemShut {NoStop}%
\bibitem [{\citenamefont {Jahn\'{a}tek}\ \emph {et~al.}(2009)\citenamefont
  {Jahn\'{a}tek}, \citenamefont {Hafner},\ and\ \citenamefont
  {Kraj\u{c}\'{\i}}}]{PhysRevB.79.224103}%
  \BibitemOpen
  \bibfield  {author} {\bibinfo {author} {\bibfnamefont {M.}~\bibnamefont
  {Jahn\'{a}tek}}, \bibinfo {author} {\bibfnamefont {J.}~\bibnamefont
  {Hafner}}, \ and\ \bibinfo {author} {\bibfnamefont {M.}~\bibnamefont
  {Kraj\u{c}\'{\i}}},\ }\href {\doibase 10.1103/PhysRevB.79.224103} {\bibfield
  {journal} {\bibinfo  {journal} {Phys. Rev. B}\ }\textbf {\bibinfo {volume}
  {79}},\ \bibinfo {pages} {224103} (\bibinfo {year} {2009})}\BibitemShut
  {NoStop}%
\bibitem [{\citenamefont {Ruban}\ and\ \citenamefont
  {Skriver}(1997)}]{PhysRevB.55.856}%
  \BibitemOpen
  \bibfield  {author} {\bibinfo {author} {\bibfnamefont {A.~V.}\ \bibnamefont
  {Ruban}}\ and\ \bibinfo {author} {\bibfnamefont {H.~L.}\ \bibnamefont
  {Skriver}},\ }\href {\doibase 10.1103/PhysRevB.55.856} {\bibfield  {journal}
  {\bibinfo  {journal} {Phys. Rev. B}\ }\textbf {\bibinfo {volume} {55}},\
  \bibinfo {pages} {856} (\bibinfo {year} {1997})}\BibitemShut {NoStop}%
\bibitem [{\citenamefont {Jiang}\ and\ \citenamefont
  {Gleeson}(2006)}]{JIANG2006433}%
  \BibitemOpen
  \bibfield  {author} {\bibinfo {author} {\bibfnamefont {C.}~\bibnamefont
  {Jiang}}\ and\ \bibinfo {author} {\bibfnamefont {B.}~\bibnamefont
  {Gleeson}},\ }\href {\doibase
  https://doi.org/10.1016/j.scriptamat.2006.05.016} {\bibfield  {journal}
  {\bibinfo  {journal} {Scripta Materialia}\ }\textbf {\bibinfo {volume}
  {55}},\ \bibinfo {pages} {433} (\bibinfo {year} {2006})}\BibitemShut
  {NoStop}%
\bibitem [{\citenamefont {Wu}\ and\ \citenamefont {Li}(2012)}]{WU2012436}%
  \BibitemOpen
  \bibfield  {author} {\bibinfo {author} {\bibfnamefont {Q.}~\bibnamefont
  {Wu}}\ and\ \bibinfo {author} {\bibfnamefont {S.}~\bibnamefont {Li}},\ }\href
  {\doibase https://doi.org/10.1016/j.commatsci.2011.09.016} {\bibfield
  {journal} {\bibinfo  {journal} {Computational Materials Science}\ }\textbf
  {\bibinfo {volume} {53}},\ \bibinfo {pages} {436} (\bibinfo {year}
  {2012})}\BibitemShut {NoStop}%
\bibitem [{\citenamefont {Rice}(1992)}]{RICE1992239}%
  \BibitemOpen
  \bibfield  {author} {\bibinfo {author} {\bibfnamefont {J.~R.}\ \bibnamefont
  {Rice}},\ }\href {\doibase https://doi.org/10.1016/S0022-5096(05)80012-2}
  {\bibfield  {journal} {\bibinfo  {journal} {Journal of the Mechanics and
  Physics of Solids}\ }\textbf {\bibinfo {volume} {40}},\ \bibinfo {pages}
  {239} (\bibinfo {year} {1992})}\BibitemShut {NoStop}%
\bibitem [{\citenamefont {Dodaran}\ \emph {et~al.}(2021)\citenamefont
  {Dodaran}, \citenamefont {Guo}, \citenamefont {Khonsari}, \citenamefont
  {Shamsaei},\ and\ \citenamefont {Shao}}]{DODARAN2021110326}%
  \BibitemOpen
  \bibfield  {author} {\bibinfo {author} {\bibfnamefont {M.~S.}\ \bibnamefont
  {Dodaran}}, \bibinfo {author} {\bibfnamefont {S.}~\bibnamefont {Guo}},
  \bibinfo {author} {\bibfnamefont {M.~M.}\ \bibnamefont {Khonsari}}, \bibinfo
  {author} {\bibfnamefont {N.}~\bibnamefont {Shamsaei}}, \ and\ \bibinfo
  {author} {\bibfnamefont {S.}~\bibnamefont {Shao}},\ }\href {\doibase
  https://doi.org/10.1016/j.commatsci.2021.110326} {\bibfield  {journal}
  {\bibinfo  {journal} {Computational Materials Science}\ }\textbf {\bibinfo
  {volume} {191}},\ \bibinfo {pages} {110326} (\bibinfo {year}
  {2021})}\BibitemShut {NoStop}%
\bibitem [{\citenamefont {Badura-Gergen}\ and\ \citenamefont
  {Schaefer}(1997)}]{PhysRevB.56.3032}%
  \BibitemOpen
  \bibfield  {author} {\bibinfo {author} {\bibfnamefont {K.}~\bibnamefont
  {Badura-Gergen}}\ and\ \bibinfo {author} {\bibfnamefont {H.-E.}\ \bibnamefont
  {Schaefer}},\ }\href {\doibase 10.1103/PhysRevB.56.3032} {\bibfield
  {journal} {\bibinfo  {journal} {Phys. Rev. B}\ }\textbf {\bibinfo {volume}
  {56}},\ \bibinfo {pages} {3032} (\bibinfo {year} {1997})}\BibitemShut
  {NoStop}%
\bibitem [{\citenamefont {Sluiter}\ and\ \citenamefont
  {Kawazoe}(1995)}]{PhysRevB.51.4062}%
  \BibitemOpen
  \bibfield  {author} {\bibinfo {author} {\bibfnamefont {M.~H.~F.}\
  \bibnamefont {Sluiter}}\ and\ \bibinfo {author} {\bibfnamefont
  {Y.}~\bibnamefont {Kawazoe}},\ }\href {\doibase 10.1103/PhysRevB.51.4062}
  {\bibfield  {journal} {\bibinfo  {journal} {Phys. Rev. B}\ }\textbf {\bibinfo
  {volume} {51}},\ \bibinfo {pages} {4062} (\bibinfo {year}
  {1995})}\BibitemShut {NoStop}%
\bibitem [{\citenamefont {Liu}\ \emph {et~al.}(2017)\citenamefont {Liu},
  \citenamefont {Wen}, \citenamefont {Li}, \citenamefont {Liu}, \citenamefont
  {Yan},\ and\ \citenamefont {Wang}}]{LIU2017157}%
  \BibitemOpen
  \bibfield  {author} {\bibinfo {author} {\bibfnamefont {S.}~\bibnamefont
  {Liu}}, \bibinfo {author} {\bibfnamefont {M.}~\bibnamefont {Wen}}, \bibinfo
  {author} {\bibfnamefont {Z.}~\bibnamefont {Li}}, \bibinfo {author}
  {\bibfnamefont {W.}~\bibnamefont {Liu}}, \bibinfo {author} {\bibfnamefont
  {P.}~\bibnamefont {Yan}}, \ and\ \bibinfo {author} {\bibfnamefont
  {C.}~\bibnamefont {Wang}},\ }\href {\doibase
  https://doi.org/10.1016/j.matdes.2017.05.032} {\bibfield  {journal} {\bibinfo
   {journal} {Materials \& Design}\ }\textbf {\bibinfo {volume} {130}},\
  \bibinfo {pages} {157} (\bibinfo {year} {2017})}\BibitemShut {NoStop}%
\bibitem [{\citenamefont {Zhu}\ \emph {et~al.}(2023)\citenamefont {Zhu},
  \citenamefont {Wang}, \citenamefont {Wang}, \citenamefont {Shi},
  \citenamefont {Liu}, \citenamefont {Li}, \citenamefont {Chen}, \citenamefont
  {Ma}, \citenamefont {Liu},\ and\ \citenamefont {Chen}}]{ZHU202354}%
  \BibitemOpen
  \bibfield  {author} {\bibinfo {author} {\bibfnamefont {H.}~\bibnamefont
  {Zhu}}, \bibinfo {author} {\bibfnamefont {J.}~\bibnamefont {Wang}}, \bibinfo
  {author} {\bibfnamefont {L.}~\bibnamefont {Wang}}, \bibinfo {author}
  {\bibfnamefont {Y.}~\bibnamefont {Shi}}, \bibinfo {author} {\bibfnamefont
  {M.}~\bibnamefont {Liu}}, \bibinfo {author} {\bibfnamefont {J.}~\bibnamefont
  {Li}}, \bibinfo {author} {\bibfnamefont {Y.}~\bibnamefont {Chen}}, \bibinfo
  {author} {\bibfnamefont {Y.}~\bibnamefont {Ma}}, \bibinfo {author}
  {\bibfnamefont {P.}~\bibnamefont {Liu}}, \ and\ \bibinfo {author}
  {\bibfnamefont {X.-Q.}\ \bibnamefont {Chen}},\ }\href {\doibase
  https://doi.org/10.1016/j.jmst.2022.10.010} {\bibfield  {journal} {\bibinfo
  {journal} {Journal of Materials Science \& Technology}\ }\textbf {\bibinfo
  {volume} {143}},\ \bibinfo {pages} {54} (\bibinfo {year} {2023})}\BibitemShut
  {NoStop}%
\bibitem [{\citenamefont {Maisel}\ \emph {et~al.}(2016)\citenamefont {Maisel},
  \citenamefont {H\"ofler},\ and\ \citenamefont
  {M\"uller}}]{PhysRevB.94.014116}%
  \BibitemOpen
  \bibfield  {author} {\bibinfo {author} {\bibfnamefont {S.~B.}\ \bibnamefont
  {Maisel}}, \bibinfo {author} {\bibfnamefont {M.}~\bibnamefont {H\"ofler}}, \
  and\ \bibinfo {author} {\bibfnamefont {S.}~\bibnamefont {M\"uller}},\ }\href
  {\doibase 10.1103/PhysRevB.94.014116} {\bibfield  {journal} {\bibinfo
  {journal} {Phys. Rev. B}\ }\textbf {\bibinfo {volume} {94}},\ \bibinfo
  {pages} {014116} (\bibinfo {year} {2016})}\BibitemShut {NoStop}%
\bibitem [{\citenamefont {Booth-Morrison}\ \emph {et~al.}(2008)\citenamefont
  {Booth-Morrison}, \citenamefont {Mao}, \citenamefont {Noebe},\ and\
  \citenamefont {Seidman}}]{doi10.1063/1.2956398}%
  \BibitemOpen
  \bibfield  {author} {\bibinfo {author} {\bibfnamefont {C.}~\bibnamefont
  {Booth-Morrison}}, \bibinfo {author} {\bibfnamefont {Z.}~\bibnamefont {Mao}},
  \bibinfo {author} {\bibfnamefont {R.~D.}\ \bibnamefont {Noebe}}, \ and\
  \bibinfo {author} {\bibfnamefont {D.~N.}\ \bibnamefont {Seidman}},\ }\href
  {\doibase 10.1063/1.2956398} {\bibfield  {journal} {\bibinfo  {journal}
  {Applied Physics Letters}\ }\textbf {\bibinfo {volume} {93}},\ \bibinfo
  {pages} {033103} (\bibinfo {year} {2008})}\BibitemShut {NoStop}%
\bibitem [{\citenamefont {Wei}\ \emph {et~al.}(2022)\citenamefont {Wei},
  \citenamefont {Lin}, \citenamefont {Huang}, \citenamefont {Huang},
  \citenamefont {Zhou}, \citenamefont {Zhang},\ and\ \citenamefont
  {Zhang}}]{WEI2022118336}%
  \BibitemOpen
  \bibfield  {author} {\bibinfo {author} {\bibfnamefont {B.}~\bibnamefont
  {Wei}}, \bibinfo {author} {\bibfnamefont {Y.}~\bibnamefont {Lin}}, \bibinfo
  {author} {\bibfnamefont {Z.}~\bibnamefont {Huang}}, \bibinfo {author}
  {\bibfnamefont {L.}~\bibnamefont {Huang}}, \bibinfo {author} {\bibfnamefont
  {K.}~\bibnamefont {Zhou}}, \bibinfo {author} {\bibfnamefont {L.}~\bibnamefont
  {Zhang}}, \ and\ \bibinfo {author} {\bibfnamefont {L.}~\bibnamefont
  {Zhang}},\ }\href {\doibase https://doi.org/10.1016/j.actamat.2022.118336}
  {\bibfield  {journal} {\bibinfo  {journal} {Acta Materialia}\ }\textbf
  {\bibinfo {volume} {240}},\ \bibinfo {pages} {118336} (\bibinfo {year}
  {2022})}\BibitemShut {NoStop}%
\bibitem [{\citenamefont {Walston}\ \emph {et~al.}(2005)\citenamefont
  {Walston}, \citenamefont {Cetel}, \citenamefont {MacKay}, \citenamefont
  {OHara}, \citenamefont {Duhl},\ and\ \citenamefont
  {Dreshfield}}]{10.7449/2004/Superalloys_2004_15_24}%
  \BibitemOpen
  \bibfield  {author} {\bibinfo {author} {\bibfnamefont {S.}~\bibnamefont
  {Walston}}, \bibinfo {author} {\bibfnamefont {A.}~\bibnamefont {Cetel}},
  \bibinfo {author} {\bibfnamefont {R.}~\bibnamefont {MacKay}}, \bibinfo
  {author} {\bibfnamefont {K.}~\bibnamefont {OHara}}, \bibinfo {author}
  {\bibfnamefont {D.}~\bibnamefont {Duhl}}, \ and\ \bibinfo {author}
  {\bibfnamefont {R.}~\bibnamefont {Dreshfield}},\ }in\ \href {\doibase
  10.7449/2004/Superalloys_2004_15_24} {\emph {\bibinfo {booktitle}
  {Superalloys 2004}}}\ (\bibinfo {year} {2005})\BibitemShut {NoStop}%
\bibitem [{\citenamefont {Zhang}\ \emph {et~al.}(2003)\citenamefont {Zhang},
  \citenamefont {Murakumo}, \citenamefont {Koizumi}, \citenamefont
  {Kobayashi},\ and\ \citenamefont {Harada}}]{ZHANG20035073}%
  \BibitemOpen
  \bibfield  {author} {\bibinfo {author} {\bibfnamefont {J.}~\bibnamefont
  {Zhang}}, \bibinfo {author} {\bibfnamefont {T.}~\bibnamefont {Murakumo}},
  \bibinfo {author} {\bibfnamefont {Y.}~\bibnamefont {Koizumi}}, \bibinfo
  {author} {\bibfnamefont {T.}~\bibnamefont {Kobayashi}}, \ and\ \bibinfo
  {author} {\bibfnamefont {H.}~\bibnamefont {Harada}},\ }\href {\doibase
  https://doi.org/10.1016/S1359-6454(03)00355-0} {\bibfield  {journal}
  {\bibinfo  {journal} {Acta Materialia}\ }\textbf {\bibinfo {volume} {51}},\
  \bibinfo {pages} {5073} (\bibinfo {year} {2003})}\BibitemShut {NoStop}%
\bibitem [{\citenamefont {Kobayashi}\ \emph {et~al.}(2005)\citenamefont
  {Kobayashi}, \citenamefont {Harada}, \citenamefont {Osawa},\ and\
  \citenamefont {Sato}}]{Kobayashi2005CreepSO}%
  \BibitemOpen
  \bibfield  {author} {\bibinfo {author} {\bibfnamefont {T.}~\bibnamefont
  {Kobayashi}}, \bibinfo {author} {\bibfnamefont {H.}~\bibnamefont {Harada}},
  \bibinfo {author} {\bibfnamefont {M.}~\bibnamefont {Osawa}}, \ and\ \bibinfo
  {author} {\bibfnamefont {A.}~\bibnamefont {Sato}},\ }\href@noop {} {\bibfield
   {journal} {\bibinfo  {journal} {Journal of The Japan Institute of Metals}\
  }\textbf {\bibinfo {volume} {69}},\ \bibinfo {pages} {1099} (\bibinfo {year}
  {2005})}\BibitemShut {NoStop}%
\bibitem [{\citenamefont {Yokokawa}\ \emph {et~al.}(2020)\citenamefont
  {Yokokawa}, \citenamefont {Harada}, \citenamefont {Kawagishi}, \citenamefont
  {Kobayashi}, \citenamefont {Yuyama},\ and\ \citenamefont
  {Takata}}]{10.1007/978-3-030-51834-9_12}%
  \BibitemOpen
  \bibfield  {author} {\bibinfo {author} {\bibfnamefont {T.}~\bibnamefont
  {Yokokawa}}, \bibinfo {author} {\bibfnamefont {H.}~\bibnamefont {Harada}},
  \bibinfo {author} {\bibfnamefont {K.}~\bibnamefont {Kawagishi}}, \bibinfo
  {author} {\bibfnamefont {T.}~\bibnamefont {Kobayashi}}, \bibinfo {author}
  {\bibfnamefont {M.}~\bibnamefont {Yuyama}}, \ and\ \bibinfo {author}
  {\bibfnamefont {Y.}~\bibnamefont {Takata}},\ }in\ \href@noop {} {\emph
  {\bibinfo {booktitle} {Superalloys 2020}}},\ \bibinfo {editor} {edited by\
  \bibinfo {editor} {\bibfnamefont {S.}~\bibnamefont {Tin}}, \bibinfo {editor}
  {\bibfnamefont {M.}~\bibnamefont {Hardy}}, \bibinfo {editor} {\bibfnamefont
  {J.}~\bibnamefont {Clews}}, \bibinfo {editor} {\bibfnamefont
  {J.}~\bibnamefont {Cormier}}, \bibinfo {editor} {\bibfnamefont
  {Q.}~\bibnamefont {Feng}}, \bibinfo {editor} {\bibfnamefont {J.}~\bibnamefont
  {Marcin}}, \bibinfo {editor} {\bibfnamefont {C.}~\bibnamefont {O'Brien}}, \
  and\ \bibinfo {editor} {\bibfnamefont {A.}~\bibnamefont {Suzuki}}}\ (\bibinfo
   {publisher} {Springer International Publishing},\ \bibinfo {address}
  {Cham},\ \bibinfo {year} {2020})\ pp.\ \bibinfo {pages}
  {122--130}\BibitemShut {NoStop}%
\bibitem [{\citenamefont {Kawagishi}\ \emph {et~al.}(2012)\citenamefont
  {Kawagishi}, \citenamefont {Yeh}, \citenamefont {Yokokawa}, \citenamefont
  {Kobayashi}, \citenamefont {Koizumi},\ and\ \citenamefont
  {Harada}}]{10.1002/9781118516430.ch21}%
  \BibitemOpen
  \bibfield  {author} {\bibinfo {author} {\bibfnamefont {K.}~\bibnamefont
  {Kawagishi}}, \bibinfo {author} {\bibfnamefont {A.-C.}\ \bibnamefont {Yeh}},
  \bibinfo {author} {\bibfnamefont {T.}~\bibnamefont {Yokokawa}}, \bibinfo
  {author} {\bibfnamefont {T.}~\bibnamefont {Kobayashi}}, \bibinfo {author}
  {\bibfnamefont {Y.}~\bibnamefont {Koizumi}}, \ and\ \bibinfo {author}
  {\bibfnamefont {H.}~\bibnamefont {Harada}},\ }\enquote {\bibinfo {title}
  {Development of an oxidation-resistant high-strength sixth-generation
  single-crystal superalloy tms-238},}\ in\ \href {\doibase
  https://doi.org/10.1002/9781118516430.ch21} {\emph {\bibinfo {booktitle}
  {Superalloys 2012}}}\ (\bibinfo  {publisher} {John Wiley \& Sons, Ltd},\
  \bibinfo {year} {2012})\ pp.\ \bibinfo {pages} {189--195}\BibitemShut
  {NoStop}%
\bibitem [{\citenamefont {Cao}\ \emph {et~al.}(2021)\citenamefont {Cao},
  \citenamefont {Yang}, \citenamefont {Chen}, \citenamefont {Liu},
  \citenamefont {Ma}, \citenamefont {Ding},\ and\ \citenamefont
  {Shi}}]{CAO2021260}%
  \BibitemOpen
  \bibfield  {author} {\bibinfo {author} {\bibfnamefont {S.}~\bibnamefont
  {Cao}}, \bibinfo {author} {\bibfnamefont {Y.}~\bibnamefont {Yang}}, \bibinfo
  {author} {\bibfnamefont {B.}~\bibnamefont {Chen}}, \bibinfo {author}
  {\bibfnamefont {K.}~\bibnamefont {Liu}}, \bibinfo {author} {\bibfnamefont
  {Y.}~\bibnamefont {Ma}}, \bibinfo {author} {\bibfnamefont {L.}~\bibnamefont
  {Ding}}, \ and\ \bibinfo {author} {\bibfnamefont {J.}~\bibnamefont {Shi}},\
  }\href {\doibase https://doi.org/10.1016/j.jmst.2021.01.049} {\bibfield
  {journal} {\bibinfo  {journal} {Journal of Materials Science \& Technology}\
  }\textbf {\bibinfo {volume} {86}},\ \bibinfo {pages} {260} (\bibinfo {year}
  {2021})}\BibitemShut {NoStop}%
\end{thebibliography}%

\end{document}